\documentclass[12pt,oneside,english]{article}
 \usepackage{epsfig}
\usepackage[T1]{fontenc}
\usepackage[latin1]{inputenc}
\usepackage{babel}
\usepackage{setspace}
\usepackage{rotating}

\makeatletter

\begin{document}

\centerline{ \bf Gravitational holography and trapped surfaces}

\bigskip \bigskip

\centerline{\large  D.Bar} 

\bigskip

\begin{abstract}
{\it  We have previously discussed the characteristics  of the
gravitational waves (GW) and have, theoretically, shown that, like the
corresponding electromagnetic (EM) waves, they also demonstrate, under certain
conditions, holographic properties. In this work we have expanded this
discussion and show that the assumed gravitational holographic images  may, 
theoretically, 
be related  to another property
of GW's which is their possible relation  to 
singular (or nonsingular) trapped surfaces.  
 We also show that this possibility may be, theoretically, related even to
weak GW's. }

\end{abstract}

 \noindent    \underline{Pacs Numbers}: 42.40.-i, 04.20.Gz, 04.30.-w, 04.30.Nk  
     \bigskip  
     
\markright{INTRODUCTION}      
\protect \section{\bf INTRODUCTION}

It is accepted in the literature that no gravitational wave (GW)
{\cite{mtw,thorne1,thorne2} passes a
spacetime region without leaving its fingerprints in this region
\cite{brill1,eppley,tipler,yurtsever1,beig,abrahams,alcubierre}.   
 As emphasized
in \cite{yurtsever1} each GW is characterized by its own intrinsic 
spacetime 
the geometry of which is imprinted  upon the passed region 
in the sense that its geometry assume the same form as that of the GW. 
  The imprinted geometry
may be either stable for a long time if the relevant GW is strong 
or transient if it  is weak
\cite{eppley,beig,abrahams,alcubierre}. This
geometry  is, theoretically, traced and  located  in the related trapped
surface \cite{brill1,eppley,abrahams,alcubierre}.  Among the
different kinds of these surfaces one may find either the  
singular trapped ones \cite{eppley,tipler,yurtsever1,abrahams} or  the 
nonsingular ones
\cite{beig} which are 
related to the  regular and 
asymptotically
flat initial data \cite{mtw,eppley,brill2,adm,nakamura} in vacuum. 
Both kinds of these surfaces  are  
  related to strong GW's where, as mentioned,  weak ones have only a
  transient influence upon spacetime. One may  note the
extensive numerical work regarding, especially,  the  collapse of  
 source-free GW's in vacuum 
\cite{eppley,abrahams,alcubierre,brill2,anninos}  to  black holes 
with the accompanying
apparent horizons \cite{eppley,abrahams,alcubierre,hawking}.  \par  
In \cite{holo} we have compared the EM theory with the linearized version of
general relativity and have arrived at a possible theory of gravitational wave
holography in which the subject, the reference \cite{gabor,collier} and the 
reconstructing \cite{gabor,collier} waves are
all GW's. We note that a more thorough comparison between the EM theory and the
linearized version  of general relativity has led Kuchar \cite{kuchar2}
to the concept of extrinsic time which is canonically conjugate to spatial 
coordinates and
not to the energy as is the usual intrinsic time.   
The basis of the comparison in
\cite{holo} which leads to the assumed holographic properties for GW's is   
  the
phase difference and the interference which are found to exist for both 
  EM \cite{gabor,collier} and gravitational \cite{holo} 
  plane waves.    This  
  may be related to the  conclusions  
  in  \cite{tipler,yurtsever1} 
that the collision (corresponds to interference) between two plane waves 
results with a  
strengthening (corresponds to constructive interference) of
them with the consequence of forming a singularity in the related region which
is, generally,  surrounded by a surface. Thus, 
the gravitational holographic image which is characterized by: \par 
(1) The same  spacetime geometry as that of the generating GW \cite{holo}.   
(2)    Changing the    spacetime
curvature in the  region  containing it relative  to neighbouring regions. 
   And 
(3) is formed from the constructive interference of two plane waves corresponds
to the trapped surface \cite{eppley,hawking} which, likewise, (1) has the same
spacetime geometry as that of the forming GW
\cite{eppley,alcubierre}.   (2) denotes a change  in spacetime
curvature in the  region which contains it \cite{brill1,eppley}. And (3) 
results also 
from the strengthening collision of two plane waves \cite{tipler,yurtsever1}. 
This  correspondence may lead one to adopt for 
 gravitational holography the same conclusions and 
   methods applied
for trapped surfaces, such as, for example,  calculating their intrinsic 
geometry.    
 \par  
We note  that although  we refer here, as  the basis for any linearized 
discussion of general relativity,  to weak GW's, which
entails only a small transient change  in spacetime
curvature   \cite{alcubierre} this small change 
may, however,  persists  if the GW which gives 
rise to it 
 stays in the related region.     
This  is the same as the    
 optics holograpic images  which are  seen only when the
reconstructing (reference) wave is sent through the hologram. That is,
 there exists a strong correspondence between the (visualized)  material of 
 the optics holographic image  which is  made from  
the light
of the  EM wave  and  the material  of the gravitational
holographic image (trapped
surface) which is related to the mass of the GW. 
Note that this mass  were shown to be 
 real and positive  even for a  source-free GW in vacuum
\cite{brill1,eppley} and  without  it 
 no change  in curvature results.  
\par 
The mentioned trapped surfaces
may, in principle, be visualized  by drawing embedding diagrams
\cite{mtw,brill1,eppley,alcubierre} of their geometry. As noted in \cite{eppley}
  it is a 
 difficult task to embed the whole trapped 
surface but one can manage to
embed the equatorial plane especially if it has rotational symmetry such as 
the  initial data of Brill \cite{eppley,alcubierre,brill2} or the Kuchar's 
cylindrical ones \cite{kuchar2}. Some embedding methods may be found, for
example, in \cite{brill2} or \cite{eppley}.  We show here that the gravitational
holographic  image, which (1) result from the constructive interference of plane
GW's \cite{holo} and (2) are restricted to small regions \cite{holo},  
may be regarded as a kind of trapped surface \cite{brill1,eppley}. For this
purpose we have calculated, using the methods in \cite{eppley}, 
 the geometry of the equatorial plane of the related holographic  image. 
 \par 
  In Section II we calculate the
relevant expressions, such as the metrics,   polarization 
and  locations of  test particles (TP),  
  for  the linearized plane GW in the
transverse traceless (TT) gauge \cite{mtw}.   The mathematical process
\cite{mtw},  leading
to the calculated locations of TP's, is
introduced, for completness, in Appendix $A$. In Section III we calculate the
relevant intensity, exposure and transmittance for 
the GW's which constitute  the gravitational holographic
process introduced in \cite{holo}. In section IV we use the
methods in \cite{eppley} for calculating and introducing  
the appropriate embedded 
trapped surfaces \cite{eppley,brill1} related to the discussed gravitational 
holography. 
The corresponding embedding, 
resulting under certain conditions from EM waves, is calculated in
Appendix $B$.  
As mentioned, the basis for the assumed 
gravitational holography was the comparison in \cite{holo} 
between EM theory and
the linearized version of general relativity. This comparison,  beginning 
from the
initial separate waves and their interference, continuing with 
their related intensity,  exposure and  
transmittance and ending with the
corresponding trapped surfaces is demonstrated in Table 1.    
 We summarize our results with
a Concluding Remarks Section.

\markright{THE LINEARIZED PLANE GRAVITATIONAL WAVE}
\protect \section{The linearized plane gravitational wave}
       
A gravitational wave is known to be characterized \cite{mtw,thorne1} by an 
oscillating
curvature tensor which causes the immediate neighbourhood of the 
space-time region through which it
pass to correspondingly oscillate \cite{mtw}.   
An example of such an oscillation is shown
in Figure 1 where a circular  array of TP's changes its form to an
elliptic  one. If the relevant region includes the location 
 of any  two neighbouring TP's, which move along geodesic lines, 
 then the interval between these
lines  
also oscillates.   We discuss two representative TP's, 
 denoted  ${\cal A}$ and ${\cal B}$, each  traversing its own  
geodesic  denoted also by the same  ${\cal A}$ and ${\cal B}$.  
  The
separation between ${\cal A}$ and ${\cal B}$ is  denoted by the vector 
${\bf n}$. 
The calculations here, such as the proper locations of ${\cal B}$ are done with 
respect to  the proper reference frame 
of ${\cal A}$.   That is, the spatial origin $x^j=0$ is located on the
world line of ${\cal A}$ and the coordinate time $x^0$ is identical to its proper 
time, e.g.,  $x^0=\tau_A$. The system is also assumed to be nonrotating   
and so it may be considered to be  a local Lorentz frame \cite{mtw,bergmann}
along the whole world line of ${\cal A}$ and not just at one event  of it. As
mentioned, we discuss
here  weak GW's and the corresponding   linearized theory of gravitation. 
Thus, one may write the
metric tensor  as 
\begin{equation} g_{\mu \nu}=\eta_{\mu \nu}+h_{\mu \nu}+O(h_{\mu \nu})^2, 
\label{e1} \end{equation} 
  where $\eta_{\mu \nu}$ is the Lorentz metric tensor \cite{mtw,bergmann} 
  and   $h_{\mu \nu}$ 
   is a small  metric perturbation which, as emphasized in \cite{mtw}, 
    denotes  also  the passing 
   GW which
is itself a traveling  perturbation in  spacetime
\cite{mtw,thorne1}. 
    The  metric, therefore,  is \cite{mtw}
\begin{equation}
ds^2=-dx^{0^2}+\delta_{jk}dx^jdx^k+O(|x^j|^2)dx^{\alpha}dx^{\beta} \label{e2} 
\end{equation} 
 We use  the transverse-traceless (TT) 
gauge \cite{mtw,thorne1} in which the tensor $h_{\mu \nu}$ is considerably 
simplified and reduces to a minimum
number of components \cite{mtw}.  In this gauge the  components of
the metric tensor satisfy (1)
$h^{TT}_{\mu 0}=0$, that is,  any component of
the metric tensor, except the spatial ones,    vanishes,  
(2)  $h^{TT}_{kj,j}=0$, so that  these components are
divergence-free and (3) are also trace-free, e.g.,   
$h^{TT}_{kk}=0$. Thus, since, as mentioned, 
the gravitational wave is the same as  
$h_{jk}^{TT}$ it, naturally,  has the same properties. \par
   The relevant  calculations for 
the changed location (due to the passing GW)  of ${\cal B}$ relative 
to ${\cal A}$ 
are introduced in
\cite{mtw} and repeated, for completness, in Appendix $A$.  The
expression for this location is  
\begin{equation} x^j_{\cal B}(\tau)=x_{{\cal B}(0)}^k(\delta_{jk}+\frac{1}{2}h^{TT}_{jk})_{at
{\cal A}} \label{e3} \end{equation}  
In the following we refer to  a plane monochromatic gravitational 
wave advancing in 
the general ${\hat{\bf
n}}$
direction where the TP's ${\cal A}$ and ${\cal B}$ and the geodesics along 
 which they propagate
 lie in the plane perpendicular to
${\hat{\bf n}}$.  Denoting the two perpendicular directions 
to ${\hat{\bf n}}$ by 
${\bf e}_{{\hat{\bf n}}_1}$, ${\bf e}_{{\hat{\bf n}}_2}$ and comparing  
the polarization
 of the GW to that of the EM waves one may realize \cite{mtw} that to 
the unit  polarization
vectors  ${\bf e}_{{\hat{\bf n}}_1}$ and ${\bf e}_{{\hat{\bf n}}_2}$ of the 
electromagnetic 
 linearly polarized wave, which propagates in the ${\hat{\bf n}}$ direction, 
  there correspond the following gravitational unit
 linear-polarization tensors  
 
 \begin{eqnarray} 
&& {\bf e}_{+_{\hat{{\bf n}}_1\hat{{\bf n}}_1}}= 
 {\bf e}_{\hat{{\bf n}}_1} \otimes {\bf e}_{\hat{{\bf n}}_1}- {\bf e}_{\hat{{\bf
 n}}_2} \otimes {\bf e}_{\hat{{\bf n}}_2}= -\biggl({\bf e}_{\hat{{\bf n}}_2} \otimes
 {\bf e}_{\hat{{\bf n}}_2}- {\bf e}_{\hat{{\bf
 n}}_1} \otimes {\bf e}_{\hat{{\bf n}}_1}\biggr)=-{\bf e}_{+_{\hat{{\bf n}}_2\hat{{\bf
 n}}_2}} \label{e4} \\ 
 && {\bf e}_{\times_{\hat{{\bf n}}_1\hat{{\bf n}}_2}}= 
 {\bf e}_{\hat{{\bf n}}_1} \otimes {\bf e}_{\hat{{\bf n}}_2}+ {\bf e}_{\hat{{\bf
 n}}_2} \otimes {\bf e}_{\hat{{\bf n}}_1}= 
\biggl( {\bf e}_{\hat{{\bf n}}_2} \otimes {\bf e}_{\hat{{\bf n}}_1}+ {\bf e}_{\hat{{\bf
 n}}_1} \otimes {\bf e}_{\hat{{\bf n}}_2}\biggr)= {\bf e}_{\times_{\hat{{\bf
 n}}_2\hat{{\bf n}}_1}},  \nonumber 
 \end{eqnarray}  
where $\otimes$ denotes the tensor product. In the following we 
 denote  by ${\bf r}$ the
position vector of a point in space and its components by ${\bf r}_1, \ {\bf
r}_2, \ {\bf r}_3$.  Thus,  taking into account, as mentioned after
Eq (\ref{e2}),  that in the $TT$
gauge the following constraints hold 
 $h^{TT}_{\mu 0}=0$,  $h^{TT}_{ij,j}=0$,  $h^{TT}_{kk}=0$  
 one realizes, using Eq (\ref{e4}),  that,  for the GW propagating
 in the  $\hat{{\bf n}}$ direction, the only nonzero components which remain 
  are \cite{mtw} 
\begin{eqnarray} && h^{TT}_{+_{{\hat{{\bf n}}_1\hat{{\bf n}}_1}}} =\Re 
\biggl( A_+{\bf e}_{+_{\hat{{\bf n}}_1\hat{{\bf n}}_1}}e^{-ift}
e^{ik{\bf r}\hat{{\bf n}}} \biggr) = 
 A_+{\bf e}_{+_{\hat{{\bf n}}_1\hat{{\bf
n}}_1}} \cdot  
\cos\biggl(k\Big({\bf r}_1\cos(\alpha)+{\bf r}_2\cos(\beta)+{\bf r}_3\cos(\eta)\Bigr)- \nonumber \\ 
&& -ft\biggr) = -h^{TT}_{+_{{\hat{{\bf n}}_2\hat{{\bf n}}_2}}}=-\Re 
\biggl( A_+{\bf e}_{+_{\hat{{\bf n}}_2\hat{{\bf n}}_2}}e^{-ift}
e^{ik{\bf r}\hat{{\bf n}}} \biggr) = \label{e5} \\ 
&& = - A_+{\bf e}_{+_{\hat{{\bf n}}_2\hat{{\bf
n}}_2}} \cdot  
\cos\biggl(k\Big({\bf r}_1\cos(\alpha)+{\bf r}_2\cos(\beta)+{\bf r}_3\cos(\eta)\Bigr)-ft\biggr)  \nonumber
\end{eqnarray} 
\begin{eqnarray} && h^{TT}_{\times_{{\hat{{\bf n}}_1\hat{{\bf n}}_2}}}=\Re 
\biggl( A_{\times}{\bf e}_{\times_{\hat{{\bf n}}_1\hat{{\bf n}}_2}}e^{-ift}
e^{ik{\bf r}\hat{{\bf n}}} \biggr) = 
 A_{\times}{\bf e}_{\times_{\hat{{\bf n}}_1\hat{{\bf
n}}_2}} \cdot  
\cos\biggl(k\Bigl({\bf r}_1\cos(\alpha)+{\bf r}_2\cos(\beta)+{\bf r}_3\cos(\eta)\Bigr)- \nonumber \\ 
&& - ft\biggr) 
= h^{TT}_{\times_{{\hat{{\bf n}}_2\hat{{\bf n}}_1}}}=\Re 
\biggl( A_{\times}{\bf e}_{\times_{\hat{{\bf n}}_2\hat{{\bf n}}_1}}e^{-ift}
e^{ik{\bf r}\hat{{\bf n}}} \biggr) = \label{e6} \\ && = 
 A_{\times}{\bf e}_{\times_{\hat{{\bf n}}_2\hat{{\bf
n}}_1}} \cdot  
\cos\biggl(k\Bigl({\bf r}_1\cos(\alpha)+{\bf r}_2\cos(\beta)+{\bf r}_3\cos(\eta)\Bigr)-ft\biggr) 
\nonumber
 \end{eqnarray} 
where $\Re$ denotes the real parts  of the expressions which follow.  
The amplitudes  $A_+$
and $A_{\times}$ are  related,  respectively,  to  the two independent modes of
polarization ${\bf e}_{+_{\hat{{\bf n}}_1\hat{{\bf n}}_1}}$  
($=-{\bf e}_{+_{\hat{{\bf n}}_2\hat{{\bf n}}_2}}$)   and 
${\bf e}_{\times_{\hat{{\bf n}}_1\hat{{\bf n}}_2}}$ 
($={\bf e}_{\times_{\hat{{\bf n}}_2\hat{{\bf n}}_1}}$), $k$ is
$\frac{2\pi}{\lambda}$, $f$ is the time frequency,  and $\cos(\alpha),\ \cos(\beta), \ \cos(\eta)$ are the
direction cosines of  $\hat{{\bf n}}$.  Thus, one may write the  perturbation 
$h^{TT}_{jk}$, 
resulting from the passing GW,  as  
\begin{eqnarray} && h^{TT}_{jk}= h^{TT}_{+_{jk}}+
h^{TT}_{\times_{jk}}=\Re 
\biggl( (A_+{\bf e}_{+_{jk}}+
A_{\times}{\bf e}_{\times_{jk}})e^{-ift}e^{ik{\bf
r}\hat{{\bf n}}}\biggr) \label{e7} \\ && = (A_+{\bf e}_{+_{jk}}+
A_{\times}{\bf e}_{\times_{jk}})
\cos\biggl(k\Bigl({\bf r}_1\cos(\alpha)+{\bf r}_2\cos(\beta)+{\bf r}_3\cos(\eta)\Bigr)-ft\biggr) 
\nonumber \end{eqnarray}   
The effect of the perturbation upon the interval between two TP's may  best be
understood by considering a large number of TP's ${\cal B}$ which form a 
closed ring
around the TP ${\cal A}$ in the center. The effect of the passing wave with either
${\bf e}_+$  or ${\bf e}_{\times}$ polarization 
 upon the ring is shown in Figure 1.  From this
figure one may realize that circular   array of the TP's ${\cal B}$ is
periodically changed by the passing plane GW to elliptic one. 
 These
periodic changes depend  upon the phase of the GW as shown in the figure. Since
we discuss here several different GW's such as the subject and the reference 
ones we denote  these waves by the appropriate suffixes $S$ (for
subject) and $R$ (for reference).    
Substituting 
from Eq (\ref{e7}) into  Eq
$(A_{11})$ of Appendix A,  which gives the 
change in the spatial interval between the TP's ${\cal A}$ and ${\cal B}$ due 
to the passing subject GW,  
 one obtains for $j={\bf n}_1$
   \begin{eqnarray} && x^{{\hat {\bf n_1}}}_{{\cal B}_S}=
   \Biggl\{x_{{\cal B}(0)_S}^{{\hat {\bf n_1}}}+\frac{1}{2}\biggl( 
   A_+{\bf e}_{+_{\hat{{\bf n}}_1\hat{{\bf n}}_1}}x_{{\cal B}(0)_S}^{{\hat {\bf n_1}}}+
A_{\times}{\bf e}_{\times_{\hat{{\bf n}}_1\hat{{\bf n}}_2}}x_{{\cal B}(0)_S}^{{\hat {\bf
n_2}}}\biggr) \cdot \label{e8} 
\\ && \cdot 
\cos\biggl( k\biggl(x\cos(\alpha)+y\cos(\beta)+z\cos(\eta)\biggr)-ft\biggr)
\Biggr\}_{at {\cal A}}  
\nonumber \end{eqnarray}  
And for $j={\bf n}_2$  one obtains 
   \begin{eqnarray} && x^{{\hat {\bf n_2}}}_{{\cal B}_S}=
   \Biggl\{x_{{\cal B}(0)_{S}}^{{\hat {\bf n_2}}}+\frac{1}{2}\biggl( 
   A_{\times}{\bf e}_{\times_{\hat{{\bf n}}_2\hat{{\bf n}}_1}}x_{{\cal B}(0)_S}^{{\hat {\bf n_1}}}+
A_+{\bf e}_{+_{\hat{{\bf n}}_2\hat{{\bf n}}_2}}x_{{\cal B}(0)_S}^{{\hat {\bf
n_2}}} \biggr) \cdot \label{e9} 
\\ && \cdot 
\cos\biggl( k\Bigl(x\cos(\alpha)+y\cos(\beta)+z\cos(\eta)\Bigr)-ft\biggr)
\Biggr\}_{at {\cal A}}  
\nonumber \end{eqnarray}  
Denoting the cosine expression $\cos\biggl(
k\Bigl(x\cos(\alpha)+y\cos(\beta)+z\cos(\eta)\Bigr)-ft\biggr)$ 
by $D$  one may realize from the last two equations 
that the location of the TP's 
${\cal B}$ has been changed, due to the passage of the GW, from the initial
value of $(x^{{\hat {\bf n_1}}}_{B(0)_S}+x_{B(0)_S}^{{\hat {\bf n_2}}})$ 
to the final one
of \begin{eqnarray} && x^{{\hat {\bf n}}}_{{\cal B}_S}= 
x^{{\hat {\bf n_1}}}_{{\cal B}(0)_S}\left( 1+\frac{D}{2}(A_+{\bf e}_{+_{\hat{{\bf n}}_1\hat{{\bf n}}_1}}+
A_{\times}{\bf e}_{\times_{\hat{{\bf n}}_2{\hat{\bf n}}_1}}) \right)+
x^{{\hat {\bf n_2}}}_{{\cal B}(0)_S}\biggl( 1+\frac{D}{2}(A_+{\bf e}_{+_{\hat{{\bf n}}_2\hat{{\bf n}}_2}}+
 \label{e10} \\ && + A_{\times}{\bf e}_{\times_{\hat{{\bf n}}_1{\hat{\bf
 n}}_2}})\biggr)=  
 (x^{{\hat {\bf n_1}}}_{{\cal B}(0)_S}+x^{{\hat {\bf
n_2}}}_{{\cal B}(0)_S})\left( 1+\frac{DA_{\times}{\bf e}_{\times_{\hat{{\bf n}}_1\hat{{\bf
n}}_2}}}{2}\right)+(x^{{\hat {\bf n_1}}}_{{\cal B}(0)_S}-x^{{\hat {\bf
n_2}}}_{{\cal B}(0)_S})\cdot \frac{DA_+{\bf e}_{+_{\hat{{\bf n}}_1\hat{{\bf
n}}_1}}}{2},
\nonumber
\end{eqnarray}
where the last result was obtained from the first of Eqs (\ref{e4}). 
One may  see from Eq (\ref{e10})  that the change in the location 
of ${\cal B}$ due to the
subject GW amounts to 
\begin{equation} \Delta x^{{\hat {\bf n}}}_{{\cal B}_S}=
(x^{{\hat {\bf n_1}}}_{{\cal B}(0)_S}+x^{{\hat {\bf
n_2}}}_{{\cal B}(0)_S})\frac{DA_{\times}{\bf e}_{\times_{\hat{{\bf n}}_1\hat{{\bf
n}}_2}}}{2}+(x^{{\hat {\bf n_1}}}_{{\cal B}(0)_S}-x^{{\hat {\bf
n_2}}}_{{\cal B}(0)_S})\cdot \frac{DA_+{\bf e}_{+_{\hat{{\bf n}}_1\hat{{\bf n}}_1}}}{2} 
\label{e11}  
\end{equation}
As shown \cite{holo} this change is added to a similar change due to a second GW
which 
 constructively interfere with the former  GW. The resulting sum is imprinted
 in spacetime in the sense that a circular  array of TP's is changed
 to an elliptic  one. The
theory of this interference and the resulting gravitational holographic  image 
has been
detaily described in \cite{holo} and  we introduce in the following section 
 the results obtained
there. 

\markright{THE GRAVITATIONAL INTENSITY, EXPOSURE......}
\protect  \section{The gravitational intensity, exposure and transmittance}
 
The reference wave denoted by the suffix $R$ may be given, as done in
\cite{holo},  by an expression
similar to Eq (\ref{e7}) and the change in the location of the TP ${\cal B}$ 
due to its passage may, likewise, be written in a form similar to Eqs 
(\ref{e8})-({\ref{e11}).  Thus, using Eq (\ref{e7}) and the expression
\cite{holo} for
the intensity of GW,   one may write the intensity of the total wave
resulting from the interference of the subject and reference waves as 
\begin{eqnarray} && I_{(S+R)_{jk}}=(h^{TT}_{+_{S_{jk}}}+
h^{TT}_{\times_{S_{jk}}})(h^{TT}_{+_{S_{jk}}}+
h^{TT}_{\times_{S_{jk}}})^{*} +(h^{TT}_{+_{R_{jk}}}+
h^{TT}_{\times_{R_{jk}}}) \cdot \nonumber \\ && 
\cdot (h^{TT}_{+_{R_{jk}}}+
h^{TT}_{\times_{R_{jk}}})^{*}  + 
<(h^{TT}_{+_{S_{jk}}}+
h^{TT}_{\times_{S_{jk}}})(h^{TT}_{+_{R_{jk}}} +
h^{TT}_{\times_{R_{jk}}})^{*} + 
\label{e12} \\ && +(h^{TT}_{+_{S_{jk}}}+
h^{TT}_{\times_{S_{jk}}})^{*}(h^{TT}_{+_{R_{jk}}} +
h^{TT}_{\times_{R_{jk}}})>, \nonumber
\end{eqnarray} 
where the asteric denotes the conjugate part of the relevant complex 
expressions. 
In the following we  append to $A_+, \ A_{\times}, \  
{\bf e}_+, \ {\bf e}_{\times}, \ {\bf r}, \ {\bf n}$ the suffixes 
$S$ and $R$ to differentiate between the subject and
reference GW's.    Thus, we use \cite{mtw,holo} for the subject and
reference GW's the following expressions;    $h^{TT}_{+_{S_{jk}}} = 
 A_{+_S}{\bf e}_{+_{S_{jk}}}e^{-if_St}
e^{ik{\bf r}_S{\hat{\bf n}}_S}, \ h^{TT}_{\times_{S_{jk}}} = 
 A_{\times_S}{\bf e}_{\times_{S_{jk}}}e^{-if_St}
e^{ik{\bf r}_S{\hat{\bf n}}_S}, \ h^{TT}_{+_{R_{jk}}} = 
 A_{+_R}{\bf e}_{+_{R_{jk}}}e^{-if_Rt}
e^{ik{\bf r}_R{\hat{\bf n}}_R}, \ h^{TT}_{\times_{R_{jk}}} = 
 A_{\times_R}{\bf e}_{\times_{R_{jk}}}e^{-if_Rt}
e^{ik{\bf r}_R{\hat{\bf n}}_R} $. Assuming, as generally done in 
optics holography, 
   the same  frequency $f_S=f_R=f$ for 
 the subject and reference GW's   one obtains for    the average 
  factors in the third 
 average term in Eq
(\ref{e12}) 
$\lim_{T \to \infty}\frac{1}{2T}\int_{-T}^{T}e^{i(f_R-f_S)t}dt=
\lim_{T \to \infty}\frac{1}{2T} \int_{-T}^{T}dt=1$. 
Thus, one may write  Eq (\ref{e12}) as 
\begin{eqnarray} && I_{(S+R)_{jk}}= \biggl(A_{+_S}{\bf e}_{+_{S_{jk}}}
 + A_{\times_S}{\bf e}_{\times_{S_{jk}}}\biggr)
  \cdot
 \biggl(A_{+_S}{\bf e}_{+_{S_{jk}}}
 + A_{\times_S}{\bf e}_{\times_{S_{jk}}} \biggr)^{*} + 
 \label{e13}
 \\ &&  
 + \biggl( A_{+_R}{\bf e}_{+_{R_{jk}}}
 + A_{\times_R}{\bf e}_{\times_{R_{jk}}} \biggr) 
 \cdot
 \biggl( A_{+_R}{\bf e}_{+_{R_{jk}}}
 + A_{\times_R}{\bf e}_{\times_{R_{jk}}} \biggr)^{*} + 
 \nonumber \\ && + 2\biggl( A_{+_S}{\bf e}_{+_{S_{jk}}} 
 + A_{\times_S}{\bf e}_{\times_{S_{jk}}} \biggr) \cdot 
 \biggl(A_{+_R}{\bf e}_{+_{R_{jk}}} 
 + A_{\times_R}{\bf e}_{\times_{R_{jk}}} \biggr)
 \cos\biggl(ik({\bf r}_S{\bf n}_S-{\bf r}_R{\bf n}_R)\biggr), 
\nonumber 
\end{eqnarray}

where the trigonometric identity 
$e^{ik({\bf r}_S{\bf n}_S-{\bf r}_R{\bf n}_R)}+
e^{-ik({\bf r}_S{\bf n}_S-{\bf r}_R{\bf n}_R)}=
2\cos\biggl(k({\bf r}_S{\bf n}_S-{\bf r}_R{\bf n}_R)\biggr)$ is used. 
One may realize from Eq (\ref{e13}) 
 that for 
$\cos\biggl(k({\bf r}_R{\bf n}_R-{\bf r}_S{\bf n}_S)\biggr)=0$  there is no
interference at all between the GW's   $h^{TT}_{+_{S_{jk}}}, \ 
  h^{TT}_{\times_{S_{jk}}}$ and  
 $h^{TT}_{+_{R_{jk}}},  \
  h^{TT}_{\times_{R_{jk}}}$   and the total 
intensity $I_{(S+R)_{jk}}$ is the addition of the separate intensities 
 $I_{S_{jk}}$ and
$I_{R_{jk}}$. That is, for 
$ (k({\bf r}_S{\bf n}_S-{\bf r}_R{\bf n}_R)\biggr)= \frac{N\pi}{2}$
where $N$ are the uneven numbers $N=1, 3, 5.....$ one obtains 
\begin{eqnarray} 
&&  I_{(S+R)_{jk_{(\cos(k({\bf r}_S{\bf n}_S-{\bf r}_R{\bf n}_R))=0)}}}=
\biggl( A_{+_S}{\bf e}_{+_{S_{jk}}}
 + A_{\times_S}{\bf e}_{\times_{S_{jk}}} \biggr) \cdot
 \biggl(A_{+_S}{\bf e}_{+_{S_{jk}}}  
  + A_{\times_S}{\bf e}_{\times_{S_{jk}}} \biggr)^{*} 
  + \nonumber \\ 
 &&  + 
 \biggl( A_{+_R}{\bf e}_{+_{R_{jk}}}
 + A_{\times_R}{\bf e}_{\times_{R_{jk}}}\biggr) \cdot  
 \biggl(A_{+_R}{\bf e}_{+_{R_{jk}}}
 + A_{\times_R}{\bf e}_{\times_{R_{jk}}} \biggr)^{*} 
 \label{e14}  
 \end{eqnarray}
For $\cos(k({\bf r}_S{\bf n}_S-{\bf r}_R{\bf n}_R)) \ne 0$ 
the interference
between the GW's $h^{TT}_{+_{S_{jk}}}, \ 
  h^{TT}_{\times_{S_{jk}}}$ and  
 $h^{TT}_{+_{R_{jk}}},  \
  h^{TT}_{\times_{R_{jk}}}$    does not vanish and the 
intensity $I_{(S+R)_{jk}}$ depends upon  the value of 
$\cos(k({\bf r}_S{\hat{\bf n}}_S-{\bf r}_R{\hat{\bf n}}_R))$.   
 Thus, for 
$\cos(k({\bf r}_R{\bf n}_R-{\bf r}_S{\bf n}_S))=\pm 1$ one  have  
\begin{eqnarray}  &&  
I_{(S+R)_{jk_{(\cos(k({\bf r}_S{\hat{\bf n}}_S-{\bf r}_R{\hat{\bf n}}_R))=\pm
1)}}}=
\biggl[\biggl( (A_{+_S}{\bf e}_{+_{S_{jk}}}
 + A_{\times_S}{\bf e}_{\times_{S_{jk}}}\biggr)
   \pm  \label{e15}  \\ && \pm 
 \biggl(A_{+_R}{\bf e}_{+_{R_{jk}}}  
  + A_{\times_R}{\bf e}_{\times_{R_{jk}}}
  \biggr)\biggr]^2,  
  \nonumber \end{eqnarray} 
  where the $+$ sign between the two terms at the right corresponds to 
  $\cos(k({\bf r}_S{\hat{\bf n}}_S-{\bf r}_R{\hat{\bf n}}_R))=1$ and the $-$
  sign to $\cos(k({\bf r}_S{\hat{\bf n}}_S-{\bf r}_R{\hat{\bf n}}_R))=-1$. 
Equation (\ref{e15})  may be
used in conjunction with Figure 1 to differentiate between constructive and
destructive interference.  We first take into account that for constructive 
interference    
the  GW's $h^{TT}_{+_{S_{jk}}}, \ h^{TT}_{+_{R_{jk}}}$ 
 and $h^{TT}_{\times_{S_{jk}}}, \   h^{TT}_{\times_{R_{jk}}}$
   must have approximately the same 
  unit polarization tensors, e. g. ,  
 ${\bf e}_{+_{S_{jk}}} \approx
 {\bf e}_{+_{R_{jk}}}, \ 
{\bf e}_{\times_{S_{jk}}} \approx
 {\bf e}_{\times_{R_{jk}}}$ and for destructive interference these GW's 
 must have approximately 
  opposite  unit 
polarization tensors, e. g. ,  
 ${\bf e}_{+_{S_{jk}}} \approx
- {\bf e}_{+_{R_{jk}}}, \ 
{\bf e}_{\times_{S_{jk}}} \approx
-{\bf e}_{\times_{R_{jk}}}$. \par Thus, one may realize that the first case of
approximate similar polarization for the GW's $h^{TT}_{+_{S_{jk}}}, \ 
 h^{TT}_{\times_{S_{jk}}}$ and  
 $h^{TT}_{+_{R_{jk}}},  \
  h^{TT}_{\times_{R_{jk}}}$ must corresponds to 
  $\cos(k({\bf r}_S{\hat{\bf n}}_S-{\bf r}_R{\hat{\bf n}}_R))
= 1$ and the second case of approximate opposite polarization for these GW's
corresponds to $\cos(k({\bf r}_S{\hat{\bf n}}_S-{\bf r}_R{\hat{\bf n}}_R))
=- 1$. This may also be realized by using the trigonometric identity  
 $\cos(k({\bf r}_S{\hat{\bf n}}_S-{\bf r}_R{\hat{\bf n}}_R))=
 \cos(k{\bf r}_S{\hat{\bf n}}_S)\cos(k{\bf r}_R{\hat{\bf n}}_R)+
  \sin(k{\bf r}_S{\hat{\bf n}}_S)\sin(k{\bf r}_R{\hat{\bf n}}_R)$  
   from which one
  may conclude that for  
  $\cos(k({\bf r}_S{\hat{\bf n}}_S-{\bf r}_R{\hat{\bf n}}_R))
= 1$ the angles $k{\bf r}_S{\hat{\bf n}}_S$ and $k{\bf r}_R{\hat{\bf n}}_R$ 
respectively related to the subject and reference GW's   
should be approximately the same or
differ by $ 2\pi n$, where $n=0, \ \pm 1,\ \pm 2, \pm 3.... $. 
The latter relations between the angles 
$k{\bf r}_S{\hat{\bf n}}_S$ and $k{\bf r}_R{\hat{\bf n}}_R$ 
 means that the 
polarizations of the  corresponding GW's  are the same and, therefore,
constructively interfere. 
In a similar manner one may realize
that for  $\cos(k({\bf r}_S{\hat{\bf n}}_S-{\bf r}_R{\hat{\bf n}}_R))
= -1$  the former  angles should be separated from each other by 
$ (2n+1)\cdot \pi$, 
where $n=0, \ \pm 1, \ \pm 2, \ \pm 3..... $ 
 which means that the corresponding
polarizations of these waves are opposite to each other and, therefore,
destructively interfere. 
Thus, since as just mentioned the two cases of 
$\cos(k({\bf r}_S{\hat{\bf n}}_S-{\bf r}_R{\hat{\bf n}}_R))
= 1$  and $\cos(k({\bf r}_S{\hat{\bf n}}_S-{\bf r}_R{\hat{\bf n}}_R))
=- 1$ respectively correspond   to the same (costructive interference)
 and opposite (destructive interference) 
polarizations for the GW's $h^{TT}_{+_{S_{jk}}}, \ 
 h^{TT}_{\times_{S_{jk}}}$ and  
 $h^{TT}_{+_{R_{jk}}},  \
  h^{TT}_{\times_{R_{jk}}}$ one may rewrite Eq
(\ref{e15})   
 as  
\begin{equation}   
I_{(S+R)_{jk_{(\cos(k({\bf r}_S{\hat{\bf n}}_S-{\bf r}_R{\hat{\bf n}}_R))
=\pm 1}}}=
\biggl( {\bf e}_{+_{S_{jk}}}
(A_{+_S}  \pm A_{+_R})
 + {\bf e}_{\times_{S_{jk}}}
 (A_{\times_S}  \pm A_{\times_R})\biggr)^2 \label{e16} 
 \end{equation}
  For further  substantiating the former discussion we refer to   
  Figure 1 and, therefore,  
   we  particularize the general indices $j$ and $k$ 
and assume, for example, $j=k={\hat{\bf n}}_1$. We also 
 denote the horizontal
and vertical axes 
 in Figure 1 by ${\hat{\bf n}}_1$ and ${\hat{\bf n}}_2$ respectively. 
  Thus, 
for $j=k={\hat{\bf n}}_1$ and 
$\cos(k({\bf r}_S{\hat{\bf n}}_S-{\bf r}_R{\hat{\bf n}}_R))=1$ Eq (\ref{e16}) 
becomes \begin{eqnarray} && I_{(S+R)_{{\hat{\bf n}}_1{\hat{\bf n}}_1(\cos(k({\bf 
r}_S{\hat{\bf n}}_S-{\bf r}_R{\hat{\bf n}}_R))
= 1)}}=
\biggl( {\bf e}_{+_{S_{{\hat{\bf n}}_1{\hat{\bf n}}_1}}}
(A_{+_S} + A_{+_R})
 \biggr)^2= \label{e17} \\ &&    =\biggl(({\bf e}_{\hat{{\bf n}}_1} \otimes {\bf e}_{\hat{{\bf n}}_1}- {\bf e}_{\hat{{\bf
 n}}_2} \otimes {\bf e}_{\hat{{\bf n}}_2})(A_{+_S} + A_{+_R})\biggr)^2,
\nonumber  \end{eqnarray}
where we use the first of Eqs (\ref{e4}). 
From the last equation one may realize that the 
 unit polarization tensors  ${\bf e}_{+_{S_{{\hat{\bf n}}_1{\hat{\bf n}}_1}}}$
 and ${\bf e}_{+_{R_{{\hat{\bf n}}_1{\hat{\bf n}}_1}}}$  
 are similar to each other which means that 
 the 
GW's $h^{TT}_{+_{R_{{\hat{\bf n}}_1{\hat{\bf n}}_1}}}$ and 
$h^{TT}_{+_{S_{{\hat{\bf n}}_1{\hat{\bf n}}_1}}}$ act in identical manner 
upon the ensemble of TP's ${\cal B}$  
which is to  turn it from a  
   circular form  to an elliptic one so that  they are 
   constructively interfering together. \par
   In a similar manner one may discuss the case of 
   $\cos(k({\bf r}_S{\hat{\bf n}}_S-{\bf r}_R{\hat{\bf n}}_R))=-1$ 
    so that  
 for $j=k={\hat{\bf n}}_1$ 
 Eq (\ref{e16}) may be rewritten as 
 \begin{eqnarray}  && I_{(S+R)_{{\hat{\bf n}}_1{\hat{\bf n}}_1(\cos(k({\bf 
r}_S{\hat{\bf n}}_S-{\bf r}_R{\hat{\bf n}}_R))
= -1)}}=
\biggl( {\bf e}_{+_{S_{{\hat{\bf n}}_1{\hat{\bf n}}_1}}}
(A_{+_S} - A_{+_R})
 \biggr)^2= \label{e18} \\ &&    =\biggl(({\bf e}_{\hat{{\bf n}}_1} \otimes {\bf e}_{\hat{{\bf n}}_1}- {\bf e}_{\hat{{\bf
 n}}_2} \otimes {\bf e}_{\hat{{\bf n}}_2})(A_{+_S} - A_{+_R})\biggr)^2,
\nonumber  \end{eqnarray}
where we again use the first of Eqs (\ref{e4}).
 From the last equation one may realize that, as mentioned, the unit
 polarization tensors  ${\bf e}_{+_{S_{{\hat{\bf n}}_1{\hat{\bf n}}_1}}}$ and 
 ${\bf e}_{+_{R_{{\hat{\bf n}}_1{\hat{\bf n}}_1}}}$ are opposite to each other. 
 That is, the 
GW's $h^{TT}_{+_{S_{{\hat{\bf n}}_1{\hat{\bf n}}_1}}}$ and 
$h^{TT}_{+_{R_{{\hat{\bf n}}_1{\hat{\bf n}}_1}}}$ act in contradicting  manner 
upon the ensemble of TP's ${\cal B}$  so that, for $A_{+_S} \approx A_{+_R}$,  
 the resulting action is almost null which
means that   they are 
   destructively interfering with each other. 
 \par
 When, however,  the general indices $jk$ are particularized to $j={\hat{\bf
 n}}_1$  
 and $k={\hat{\bf n}}_2$ then one may follow the former discussion and conclude 
 that for $\cos(k({\bf 
r}_S{\hat{\bf n}}_S-{\bf r}_R{\hat{\bf n}}_R))
= 1$
  the polarization
tensors ${\bf e}_{\times_{S_{{\hat{\bf n}}_1{\hat{\bf n}}_2}}}, \ 
{\bf e}_{\times_{R_{{\hat{\bf n}}_1{\hat{\bf n}}_2}}}$ of the respective GW's 
$h^{TT}_{\times_{S_{{\hat{\bf n}}_1{\hat{\bf n}}_2}}}, \  
h^{TT}_{\times_{R_{{\hat{\bf n}}_1{\hat{\bf n}}_2}}}$ are similar to each other. 
Thus, Eq (\ref{e16}) may be rewritten in this case as 
 \begin{eqnarray}  && I_{(S+R)_{{\hat{\bf n}}_1{\hat{\bf n}}_2(\cos(k({\bf 
r}_S{\hat{\bf n}}_S-{\bf r}_R{\hat{\bf n}}_R))
= 1)}}=
\biggl( {\bf e}_{\times_{S_{{\hat{\bf n}}_1{\hat{\bf n}}_2}}}
(A_{\times_S} + A_{\times_R})
 \biggr)^2= \label{e19} \\ &&   =\biggl(({\bf e}_{\hat{{\bf n}}_1} \otimes {\bf e}_{\hat{{\bf n}}_2}
 + {\bf e}_{\hat{{\bf
 n}}_2} \otimes {\bf e}_{\hat{{\bf n}}_1})(A_{\times_S} + 
 A_{\times_R})\biggr)^2, \nonumber \end{eqnarray}
where we use the second of Eqs (\ref{e4}). 
As seen from the last equation  the 
 unit polarization tensors  ${\bf e}_{\times_{S_{{\hat{\bf n}}_1{\hat{\bf n}}_2}}}$
 and ${\bf e}_{\times_{R_{{\hat{\bf n}}_1{\hat{\bf n}}_2}}}$  
 are similar to each other which means that 
 the 
GW's $h^{TT}_{\times_{S_{{\hat{\bf n}}_1{\hat{\bf n}}_2}}}$ and 
$h^{TT}_{\times_{R_{{\hat{\bf n}}_1{\hat{\bf n}}_2}}}$ act in identical manner 
upon the ensemble of TP's ${\cal B}$  
which, as seen from Figure 1,  is composed of; (1)   turning it from a  
   circular form  to an elliptic one and (2) rotating it by 45 degrees 
   so that  they are 
   constructively interfering together. 
  The case of   
   $\cos(k({\bf r}_S{\hat{\bf n}}_S-{\bf r}_R{\hat{\bf n}}_R))=-1$ 
   may also be discussed in a similar manner so that     
 for $j={\hat{\bf n}}_1, \ k={\hat{\bf n}}_2$ 
 Eq (\ref{e16}) may be rewritten as 
 \begin{eqnarray} && I_{(S+R)_{{\hat{\bf n}}_1{\hat{\bf n}}_2(\cos(k({\bf 
r}_S{\hat{\bf n}}_S-{\bf r}_R{\hat{\bf n}}_R))
= -1)}}=
\biggl( {\bf e}_{\times_{S_{{\hat{\bf n}}_1{\hat{\bf n}}_2}}}
(A_{\times_S} - A_{\times_R})
 \biggr)^2= \label{e20} \\ &&      
 =\biggl(({\bf e}_{\hat{{\bf n}}_1} \otimes {\bf e}_{\hat{{\bf n}}_2}+
  {\bf e}_{\hat{{\bf
 n}}_2} \otimes {\bf e}_{\hat{{\bf n}}_1})(A_{\times_S} - A_{\times_R})\biggr)^2,
\nonumber  \end{eqnarray} 
where we again use the second  of Eqs (\ref{e4}).
 As realized from the last equation  the unit
 polarization tensors  ${\bf e}_{\times_{S_{{\hat{\bf n}}_1{\hat{\bf n}}_2}}}$ and 
 ${\bf e}_{\times_{R_{{\hat{\bf n}}_1{\hat{\bf n}}_2}}}$ are opposite 
 to each other. 
 That is, the 
GW's $h^{TT}_{\times_{S_{{\hat{\bf n}}_1{\hat{\bf n}}_2}}}$ and 
$h^{TT}_{\times_{R_{{\hat{\bf n}}_1{\hat{\bf n}}_2}}}$ act in contradicting  
manner 
upon the ensemble of TP's ${\cal B}$  so that, for $A_{\times_S} \approx
A_{\times_R}$ the resulting action is almost null which
means that   they are 
   destructively interfering with each other. \par
 As for the gravitational hologram we may understand its nature, as emphasized
 in  \cite{holo}, from 
 the known corresponding
  holograms used in optics holography. The latter are prepared
  \cite{gabor,collier} so as
  to efficiently record the initial constructive interference of the subject and
  reference waves so that directing later the reference wave upon its surface
  results in reconstructing the initial subject wave \cite{gabor,collier}. \par 
   A great simplification of the recording
  process is obtained in optics holography, as done in \cite{collier}, 
  when one discusses a
  small area holograms in which case one may assume a linear recording process
  \cite{collier}. This process is  also  used in \cite{holo}  for the
  gravitational case and were theoretically shown that one may reconstruct the
  initial subject gravitational wave.  
   That is,  as shown in \cite{holo}, 
    the imprinted spacetime of the subject GW 
    $h^{TT}_{S_{jk}}=h^{TT}_{+_{S_{jk}}}+h^{TT}_{\times_{S_{jk}}}$ upon the 
    small region $A$ (see Figure 2) 
    becomes effective in the sense that  if a  
  reconstructing GW,  which
     should be identical to the original reference wave, passes through this 
       region   the  effect is to 
    cause  
    a contraction  of $A$  along some axis and
  elongation along  another as shown in Figure 1. This is the meaning by which
  gravitational holographic images should be understood. 
   In other words, assuming as in optics
      holography \cite{gabor,collier}, that the spacetime region exposed to the
      interfering subject and reference GW's acquires some transmittance ${\bf
      t}^{TT}_E$  
      which depends upon   this exposure $E$ one may  suppose the following:
      \par
      (1)  the exposure $E$ is   
        proportional to 
      the intensity $I_{S+R}$ from Eq (\ref{e13}) so that  $E=kI_{S+R}\tau_e$
     where  $\tau_E$ is the exposure time and $k$ a proportionality constant.
     \par 
      (2) the exposure $E$ 
     may be  written as a sum $E(r)=E_0+E_1(r)$  of a constant term
  $E_0$ and a space dependent one $E_1(r)$ where  the restriction 
  to the small region $A$ enables one to sustain the inequality $E_1(r) < E_0$ 
  over $A$. \par (3) Using (2) one may write the transmittance 
  ${\bf t}_E^{TT}$ over $A$ as a Taylor series 
  in which the coefficients of the second and higher order terms may be 
  neglected. That is   
  \begin{equation}  \label{e21} {\bf t}_E^{TT}={\bf t}^{TT}(E^{TT}_0)+ 
  E^{TT}_1 \frac{d{\bf
t}_E}{dE}|_{E^{TT}_0}+  \frac{1}{2}(E^{TT}_1 )^2\frac{d^2{\bf t}_E}{dE^2}|_{E^{TT}_0}+
\cdots, \end{equation} where  
$\frac{d^2{\bf t}_E}{dE^2}|_{E^{TT}_0} \approx 
\frac{d^3{\bf t}_E}{dE^3}|_{E^{TT}_0} \approx \approx ....0$.  
  Thus, as for optics hologram
  \cite{gabor,collier} and as shown in \cite{holo}, for reconstructing 
  the subject wave
  $h^{TT}_{S_{jk}}$ one should send through $A$  a reconstructing GW, which 
  is identical to the reference
  wave $h^{TT}_{R_{jk}}$, so that using (1)-(3) and Eq (\ref{e21})   one 
   obtains     
      \begin{equation} 
 W=h^{TT}_{R_{jk}}{\bf t}_E^{TT}=h^{TT}_{R_{jk}}E^{TT}_1(r)\frac{d{\bf
t}_E}{dE}|_{E^{TT}_0}
 =C_{TT}\cdot h^{TT}_{S_{jk}},  \label{e22}  
  \end{equation}
where $C_{TT}$ is a proportionality constant. In Table 1 we have outlined and
followed the
whole gravitational holographic process from the initial separate subject and
reference GW's until the final formed trapped surface. Since, as mentioned, this
process is based upon the comparison done in \cite{holo} between the optical
holographic theory \cite{gabor,collier} and the linearized version of general
relativity \cite{mtw,thorne1} we have also described in Table 1 the
corresponding optics holographic process. Thus, one may see in this table 
side by side  the corresponding expressions for the two processes. \par 
We should
note here that unlike the EM field which may, experimentally, be traced and
located in any region in space however small it is the gravitational field 
  can
not be located \cite{mtw} in such a manner. That is, as
emphasized in \cite{mtw} (see P. 955 in \cite{mtw}), {\it ``the stress-energy 
carried
by  gravitational waves can not be localized inside a wavelength etc. However,
one can say that a certain amount of stress-energy is contained in a given
"macroscopic" region''}. That is, the gravitational wave formalism is applied for
averages over several wavelengths. However, as mentioned, for very small
wavelength, which is the limit discussed here,  not only this formalism is 
 valid but also the comparison between it
and the EM one.  

\markright{CALCULATION OF THE EMBEDDED SURFACE}
\protect \section{Calculation of the embedded surface}

 As mentioned, the holographic
image over $A$ have a spacetime geometry which is the same as the spacetime
geometry of the subject GW $h^{TT}_{S_{jk}}$. We have also mentioned   
that    the spacetime geometry  of the trapped
surface is the same as that of the GW which gives rise to 
it \cite{eppley,brill2}. This similarity between the holographic images 
 and trapped surface  lead us to suppose that, theoretically, 
 they are similar entities. Thus, one may be tempted to use the known embedding 
 methods
 \cite{eppley,brill2} of  calculating the geometry of trapped surfaces for 
 finding
 the geometry of the gravitational holographic images. Now, since the embedding
 of the calculated geometry is into the Euclidean space \cite{eppley} 
 we have first to convert the
 tensor metric components $h^{TT}_{\hat{{\bf n}}_1\hat{{\bf n}}_1}= 
 -h^{TT}_{\hat{{\bf n}}_2\hat{{\bf n}}_2}$, 
 $h^{TT}_{\hat{{\bf n}}_1\hat{{\bf n}}_2}= 
 h^{TT}_{\hat{{\bf n}}_2\hat{{\bf n}}_1}$ from Eqs (\ref{e5})-(\ref{e6}),  which
 were calculated in the  
 ${\hat{\bf n}}, \ {\hat{\bf n}}_1, \   {\hat{\bf n}}_2 $ system, into the 
 ${\hat{\bf x}}, \ {\hat{\bf y}}, \  {\hat{\bf z}}$
 Euclidean system. That is, substituting ${\hat{\bf n}}={\hat{\bf z}},  \ 
 {\hat{\bf n}}_1={\hat{\bf x}}, \ {\hat{\bf n}}_2={\hat{\bf y}}, 
 \ {\bf r}= {\hat{\bf x}}x+{\hat{\bf y}}y+{\hat{\bf z}}z$ one may write 
 the Euclidean metric components as 
  \begin{eqnarray} && h^{TT}_{{\hat {\bf x}}{\hat{\bf x}}} =\Re 
\biggl( A_+{\bf e}_{+_{{\hat {\bf x}}{\hat{\bf x}}}}e^{-ift}
e^{ik{\bf r}\hat{{\bf z}}} \biggr) = 
 A_+{\bf e}_{+_{{\hat {\bf x}}{\hat{\bf x}}}} \cdot  
\cos(kz -ft) =     \nonumber \\ && = -h^{TT}_{{\hat {\bf y}}{\hat{\bf y}}}=-\Re 
\biggl( A_+{\bf e}_{+_{{\hat {\bf y}}{\hat{\bf y}}}}e^{-ift}
e^{ik{\bf r}\hat{{\bf z}}} \biggr) =  - A_+{\bf e}_{+_{{\hat {\bf y}}{\hat{\bf y}}}} \cdot  
\cos(kz-ft)  \label{e23}   \\  && h^{TT}_{{\hat {\bf x}}{\hat{\bf y}}}=\Re 
\biggl( A_{\times}{\bf e}_{\times_{{\hat {\bf x}}{\hat{\bf y}}}}e^{-ift}
e^{ik{\bf r}\hat{{\bf z}}} \biggr) = 
 A_{\times}{\bf e}_{\times_{{\hat {\bf x}}{\hat{\bf y}}}} \cdot  
\cos(kz- ft) 
=   \nonumber \\ && = h^{TT}_{{\hat {\bf y}}{\hat{\bf x}}}=\Re 
\biggl( A_{\times}{\bf e}_{\times_{{\hat {\bf y}}{\hat{\bf x}}}}e^{-ift}
e^{ik{\bf r}\hat{{\bf z}}} \biggr)  = 
 A_{\times}{\bf e}_{\times_{{\hat {\bf y}}{\hat{\bf x}}}} \cdot  
\cos(kz-ft),  
\nonumber
 \end{eqnarray}    
 where ${\bf e}_{+_{{\hat {\bf x}}{\hat{\bf x}}}}, \  
 {\bf e}_{+_{{\hat {\bf y}}{\hat{\bf y}}}}, \ 
 {\bf e}_{\times_{{\hat {\bf x}}{\hat{\bf y}}}}$ 
 are the Euclidean unit linear-polarization tensors given by 
 Eqs (\ref{e4}) in which we substitute 
 ${\hat{\bf n}}={\hat{\bf z}},  \ 
 {\hat{\bf n}}_1={\hat{\bf x}}, \ {\hat{\bf n}}_2={\hat{\bf y}}$.  Thus, using
 the last equations and the discussion after Eq (\ref{e2}) one may write the
 metric from Eq (\ref{e2}) in the TT gauge as 
 \begin{eqnarray} && (ds^{TT})_{({\hat{\bf x}},{\hat{\bf y}},
{\hat{\bf z}})}^2=h^{TT}_{{\hat {\bf x}}{\hat{\bf x}}}dx^2+
h^{TT}_{{\hat {\bf y}}{\hat{\bf y}}}dy^2+2h^{TT}_{{\hat {\bf x}}{\hat{\bf y}}}dxdy= 
 A_+{\bf e}_{+_{{\hat {\bf x}}{\hat{\bf x}}}} \cdot  
\cos(kz -ft)dx^2+ \nonumber \\ && + A_+{\bf e}_{+_{{\hat {\bf y}{\hat {\bf
y}}}}} \cdot  
\cos(kz-ft)dy^2+2A_{\times}{\bf e}_{\times_{{\hat {\bf x}}{\hat {\bf y}}}} \cdot  
\cos(kz-ft)dxdy = \label{e24}  \\ 
&& = A_+{\bf e}_{+_{{\hat {\bf x}}{\hat{\bf x}}}} \cdot  
\cos(kz -ft)(dx^2-dy^2)+ 2A_{\times}{\bf e}_{\times_{{\hat{\bf x}{\hat {\bf
y}}}}} \cdot  
\cos(kz-ft)dxdy,   \nonumber \end{eqnarray} 
where the last result was obtained by using 
${\bf e}_{+_{{\hat {\bf x}}{\hat{\bf x}}}}=-{\bf e}_{+_{{\hat {\bf y}}{\hat{\bf
y}}}}$ (see the first of Eqs (\ref{e4})). 
In order to calculate the embedded surface of the holographic image 
we first express the  metric from Eq
(\ref{e24})   in the cylindrical coordinates
$({\hat{\bf \rho}}, \ {\hat{\bf \phi}}, \ {\hat {\bf z}})$ 
where $x=\rho\cos(\phi), \ y=\rho\sin(\phi), \ z=z$  so that  
\begin{eqnarray} && (ds^{TT})_{({\hat{\bf \rho}},{\hat{\bf \phi}},
{\hat{\bf z}})}^2=
h^{TT}_{{\hat {\bf \rho}}{\hat{\bf \rho}}}d\rho^2+
h^{TT}_{{\hat {\bf \phi}}{\hat{\bf \phi}}}d\phi^2+
h^{TT}_{{\hat {\bf \rho}}{\hat{\bf \phi}}}d\rho
d\phi= \nonumber \\ && 
= A_+{\bf e}_{+_{{\hat {\bf \rho}}{\hat{\bf \rho}}}} \cdot  
\cos(kz -ft)\bigg(\cos(2\phi)(d\rho^2 -\rho^2d\phi^2)- 
2\rho \sin(2\phi)d\rho d\phi \biggr) 
\label{e25} \\ && +A_{\times}{\bf e}_{\times_{{\hat {\bf \rho}}{\hat{\bf \phi}}}} \cdot  
\cos(kz-ft)\biggl( \sin(2\phi)(d\rho^2-\rho^2d\phi^2)+
2\rho \cos(2\phi)d\rho d\phi \biggr), \nonumber  \end{eqnarray} 
where the following trigonometric relations were used
$(\cos^2(\phi)-\sin^2(\phi))=\cos(2\phi), \ 2\sin(\phi)\cos(\phi)=\sin(2\phi)$. 
We have also transformed from the cartesian unit polarization tensors 
${\bf e}_{+_{{\hat {\bf x}}{\hat {\bf x}}}}, \  {\bf e}_{+_{{\hat {\bf xy}}}}$ 
to the
corresponding cylindrical ones ${\bf e}_{+_{{\hat {\bf \rho}}{\hat {\bf \rho}}}}, \ 
{\bf e}_{+_{{\hat {\bf \rho}}{\hat {\bf \phi}}}}$ by using: \par
(1) The unit
polarization tensors in the $(  {\bf e}_{\hat{\bf x}}, \ 
{\bf e}_{\hat{\bf y}},  \ {\bf e}_{\hat{\bf z}})$ system
$ {\bf e}_{+_{{\hat{\bf x}}{\hat{\bf x}}}}= 
 {\bf e}_{\hat{\bf x}} \otimes {\bf e}_{\hat{\bf x}}- {\bf e}_{\hat{\bf
 y}} \otimes {\bf e}_{\hat{\bf y}}= -{\bf e}_{+_{{\hat{\bf y}}{\hat{\bf
 y}}}}, \ \   
  {\bf e}_{\times_{{\hat{\bf x}}{\hat{\bf y}}}}= 
 {\bf e}_{\hat{\bf x}} \otimes {\bf e}_{\hat{\bf y}}+ {\bf e}_{\hat{\bf
 y}} \otimes {\bf e}_{\hat{\bf x}}= {\bf e}_{\times_{\hat{\bf
 y}{\hat{\bf x}}}}$ (see Eqs (\ref{e4})). \par
  (2)  The transformation relations from the $({\bf e}_{\hat {x}}, 
  {\bf e}_{\hat {y}}, {\bf e}_{\hat {z}})$ coordinate system to the cylindrical
  one $({\bf e}_{\hat {\rho}}, 
  {\bf e}_{\hat {\phi}}, {\bf e}_{\hat {z}})$
\cite{spiegel}  
 $$ {\bf e}_{\hat {\bf x}}=\cos(\phi){\bf e}_{\hat {\bf \rho}}-\sin(\phi){\bf
  e}_{\hat {\phi}}, \ \ {\bf e}_{\hat {\bf y}}=
  \sin(\phi){\bf e}_{\hat {\bf \rho}}+\cos(\phi){\bf
  e}_{\hat {\phi}}, \  \ {\bf e}_{\hat {\bf z}}= {\bf e}_{\hat {\bf z}},  
 $$
and (3) the triginometric identities
$(\cos^2(\phi)-\sin^2(\phi))=\cos(2\phi), \ \ 2\cos(\phi)\sin(\phi)=\sin(2\phi)$.  
Thus, the cylindrical unit polarization tensors 
${\bf e}_{+_{{\hat {\bf \rho}}{\hat {\bf \rho}}}}, \ 
{\bf e}_{+_{{\hat {\bf \rho}}{\hat {\bf \phi}}}}$ in Eq (\ref{e25}) are, respectively,  given 
by 
\begin{eqnarray} && {\bf e}_{+_{{\hat {\bf \rho}}{\hat {\bf \rho}}}}=
 \cos(2\phi)\biggl( {\bf e}_{\hat {\bf \rho}}\otimes 
{\bf e}_{\hat {\bf \rho}}-{\bf e}_{\hat {\bf \phi}}\otimes 
{\bf e}_{\hat {\bf \phi}}\biggr)-\sin(2\phi)\biggl( {\bf e}_{\hat {\bf \rho}}\otimes 
{\bf e}_{\hat {\bf \phi}}+{\bf e}_{\hat {\bf \phi}}\otimes 
{\bf e}_{\hat {\bf \rho}}\biggr) \label{e26} \\ 
&& {\bf e}_{\times_{{\hat {\bf \rho}}{\hat {\bf \phi}}}}=
 \sin(2\phi)\biggl( {\bf e}_{\hat {\bf \rho}}\otimes 
{\bf e}_{\hat {\bf \rho}}-{\bf e}_{\hat {\bf \phi}}\otimes 
{\bf e}_{\hat {\bf \phi}}\biggr)+\cos(2\phi)\biggl( {\bf e}_{\hat {\bf \rho}}\otimes 
{\bf e}_{\hat {\bf \phi}}+{\bf e}_{\hat {\bf \phi}}\otimes 
{\bf e}_{\hat {\bf \rho}}\biggr)
 \nonumber \end{eqnarray} 
We assume here, as in \cite{brill1,eppley},  a no-rotation case 
 so that   $h^{TT}_{{\hat {\bf \rho}}{\hat{\bf \phi}}}$ is identically zero.  
Thus,  removing from Eq (\ref{e25}) the $ d\rho d\phi$ 
part one obtains 
 \begin{eqnarray} && (ds^{TT})_{({\hat{\bf \rho}},{\hat{\bf \phi}},
{\hat{\bf z}})}^2=
h^{TT}_{{\hat {\bf \rho}}{\hat{\bf \rho}}}d^2\rho +
h^{TT}_{{\hat {\bf \phi}}{\hat{\bf \phi}}}d^2\phi= 
\label{e27}  \\ && 
= \cos(kz -ft)\biggl(A_+{\bf e}_{+_{{\hat {\bf \rho}}{\hat{\bf \rho}}}}\cos(2\phi)
 +A_{\times}{\bf e}_{\times_{{\hat {\bf \rho}}{\hat{\bf
 \phi}}}}\sin(2\phi)\biggr)\biggl((d\rho^2 - \rho^2d\phi^2)\biggr) \nonumber 
 \end{eqnarray}
    We, now,  find  the embedding of the holographic image and begin by 
assuming, as
 for the trapped surfaces discussed in \cite{eppley}, that its metric in the
 small surface $A$ (see Figure 2) is 
 that of a surface of rotation $z(x,y)$ related to Euclidean space. 
 That is, one may write 
 \begin{equation} x=a(\rho)\cos(\phi), \ \ y=a(\rho)\sin(\phi), \ \ z=b(\rho) 
 \label{e28} \end{equation} 
 Thus,  using the expressions for $h^{TT}_{{\hat {\bf \rho}}{\hat{\bf \rho}}}$ 
 and $h^{TT}_{{\hat {\bf \phi}}{\hat{\bf \phi}}}$
 from Eq (\ref{e27}) one may write the metric of the holographic image as 
 \begin{eqnarray}  &&   
 ds^2=dx^2+dy^2+dz^2=(a^2(\rho)_{\rho}+b^2(\rho)_{\rho})d\rho^2+
 a^2(\rho)d\phi^2=  \label{e29} \\ && = 
 h^{TT}_{{\hat {\bf \rho}}{\hat{\bf \rho}}}d^2\rho +
h^{TT}_{{\hat {\bf \phi}}{\hat{\bf \phi}}}d^2\phi  = 
\cos(kz -ft)\biggl(A_+{\bf e}_{+_{{\hat {\bf \rho}}{\hat{\bf \rho}}}}\cos(2\phi)
  + A_{\times}{\bf e}_{\times_{{\hat {\bf \rho}}{\hat{\bf
 \phi}}}}\sin(2\phi)\biggr)\biggl(d\rho^2 - \nonumber \\ && - 
\rho^2d\phi^2\biggr)= \cos(kz -ft)\biggl[\frac{\sin(4\phi)}{2}(A_{\times}-A_+)
\left({\bf e}_{\hat {\bf \rho}}\otimes 
{\bf e}_{\hat {\bf \phi}}+{\bf e}_{\hat {\bf \phi}}\otimes 
{\bf e}_{\hat {\bf \rho}}\right)+\nonumber \\ && + 
\left(A_+\cos^2(2\phi)+A_{\times}\sin^2(2\phi)\right)
 \biggl( {\bf e}_{\hat {\bf \rho}}\otimes 
{\bf e}_{\hat {\bf
\rho}}-{\bf e}_{\hat {\bf \phi}}\otimes 
{\bf e}_{\hat {\bf
\phi}}\biggr)\biggr]
 \left(d^2\rho-\rho^2d\phi^2\right) \nonumber 
 \end{eqnarray}
 where $a_{\rho}(\rho), \ b_{\rho}(\rho)$ denote the first derivatives of $a, \ b$ with respect
 to $\rho$ and the last result is obtained by  substituting from Eq (\ref{e26}) 
 for  ${\bf e}_{\times_{{\hat {\bf \rho}}{\hat{\bf
 \phi}}}}, \ {\bf e}_{+_{{\hat {\bf \rho}}{\hat{\bf
 \rho}}}}$. Note that when the amplitudes $A_{\times}, \ A_+$ are equal  
  the expression $\biggl(A_+{\bf e}_{+_{{\hat {\bf \rho}}{\hat{\bf \rho}}}}\cos(2\phi)
  + A_{\times}{\bf e}_{\times_{{\hat {\bf \rho}}{\hat{\bf
 \phi}}}}\sin(2\phi)\biggr)$ is considerably simplified and becomes 
  $$\biggl(A_+{\bf e}_{+_{{\hat {\bf \rho}}{\hat{\bf \rho}}}}\cos(2\phi)
  + A_{\times}{\bf e}_{\times_{{\hat {\bf \rho}}{\hat{\bf
 \phi}}}}\sin(2\phi)\biggr)=A_+\biggl({\bf e}_{\hat {\bf \rho}}\otimes 
{\bf e}_{\hat {\bf \rho}}-{\bf e}_{\hat {\bf \phi}}\otimes 
{\bf e}_{\hat {\bf \phi}}\biggr)$$ 
 The quantities $a(\rho)$, $a_{\rho}(\rho)$ and $b(\rho)$ which
 defines the intrinsic geometry of the holographic image upon the small area 
 $A$
 are determined from Eq
 (\ref{e29}) as
 \begin{eqnarray} && a(\rho)= \rho\biggl[\cos(kz
 -ft)\biggl\{\frac{\sin(4\phi)}{2}(A_+-A_{\times})
\left({\bf e}_{\hat {\bf \rho}}\otimes 
{\bf e}_{\hat {\bf \phi}}+{\bf e}_{\hat {\bf \phi}}\otimes 
{\bf e}_{\hat {\bf \rho}}\right)+\nonumber \\ && + 
\left(A_+\cos^2(2\phi)+A_{\times}\sin^2(2\phi)\right)
 \biggl( {\bf e}_{\hat {\bf \phi}}\otimes 
{\bf e}_{\hat {\bf
\phi}}-{\bf e}_{\hat {\bf \rho}}\otimes 
{\bf e}_{\hat {\bf
\rho}}\biggr)\biggr\}\biggr]^{\frac{1}{2}} \nonumber  \\ && 
 a(\rho)_{\rho}= \biggl[\cos(kz
 -ft)\biggl\{\frac{\sin(4\phi)}{2}(A_+-A_{\times})
\left({\bf e}_{\hat {\bf \rho}}\otimes 
{\bf e}_{\hat {\bf \phi}}+{\bf e}_{\hat {\bf \phi}}\otimes 
{\bf e}_{\hat {\bf \rho}}\right)+\nonumber \\ && + 
\left(A_+\cos^2(2\phi)+A_{\times}\sin^2(2\phi)\right)
 \biggl( {\bf e}_{\hat {\bf \phi}}\otimes 
{\bf e}_{\hat {\bf
\phi}}-{\bf e}_{\hat {\bf \rho}}\otimes 
{\bf e}_{\hat {\bf
\rho}}\biggr)\biggl\}\biggr]^{\frac{1}{2}}\label{e30} \\ 
&& b(\rho)=\int d\rho\biggl[\cos(kz -ft)\biggl\{\frac{\sin(4\phi)}{2}(A_{\times}-A_+)
\left({\bf e}_{\hat {\bf \rho}}\otimes 
{\bf e}_{\hat {\bf \phi}}+{\bf e}_{\hat {\bf \phi}}\otimes 
{\bf e}_{\hat {\bf \rho}}\right)+\nonumber \\ && + 
\left(A_+\cos^2(2\phi)+A_{\times}\sin^2(2\phi)\right)
 \biggl( {\bf e}_{\hat {\bf \rho}}\otimes 
{\bf e}_{\hat {\bf
\rho}}-{\bf e}_{\hat {\bf \phi}}\otimes 
{\bf e}_{\hat {\bf
\phi}}\biggr)\biggr\}-a^2(\rho)_{\rho}\biggr]^{\frac{1}{2}}
\nonumber  \end{eqnarray} 
Note that the geometry of the small surface $A$ determined from the former 
quantities $a(\rho), \ a(\rho)_{\rho}, \
b(\rho)$ is  that of the  subject GW $h^{TT}_{S_{{\hat{\bf \rho}},{\hat {\bf 
\phi }},{\hat{\bf z}}}}$.

\begin{figure}
\centerline{
\epsfxsize=5 in
\begin{turn}{-90}
\epsffile{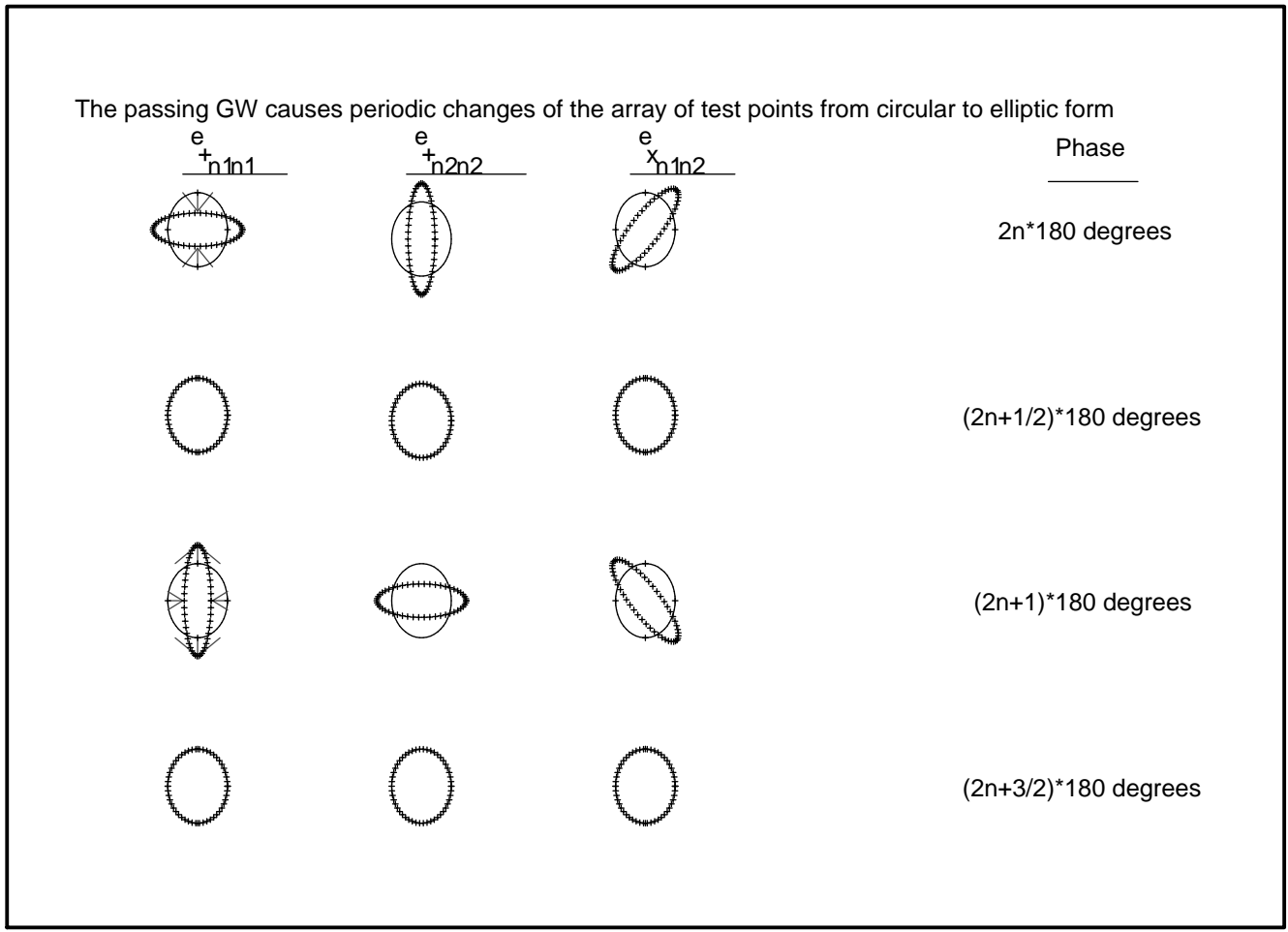}
\end{turn}}
     
    \caption{A schematic representation of the influence of a passing plane GW
upon a circular  array of test particles which periodically changes
its form to elliptic  array. The first column at the left shows the influence of
the unit polarization tensor ${\bf e}_{+_{{\hat{\bf n}}_1{\hat{\bf n}}_1}}$, 
the middle
column shows that of the unit polarization tensor 
${\bf e}_{+_{{\hat{\bf n}}_2{\hat{\bf n}}_2}}$ and the right column represents 
 ${\bf e}_{\times_{{\hat{\bf n}}_1{\hat{\bf n}}_2}}$. Note that 
 ${\bf e}_{+_{{\hat{\bf n}}_1{\hat{\bf n}}_1}}=
 -{\bf e}_{+_{{\hat{\bf n}}_2{\hat{\bf n}}_2}}$}
 
     \end{figure}

\begin{figure}
\centerline{
\epsfxsize=5 in
\begin{turn}{-90}
\epsffile{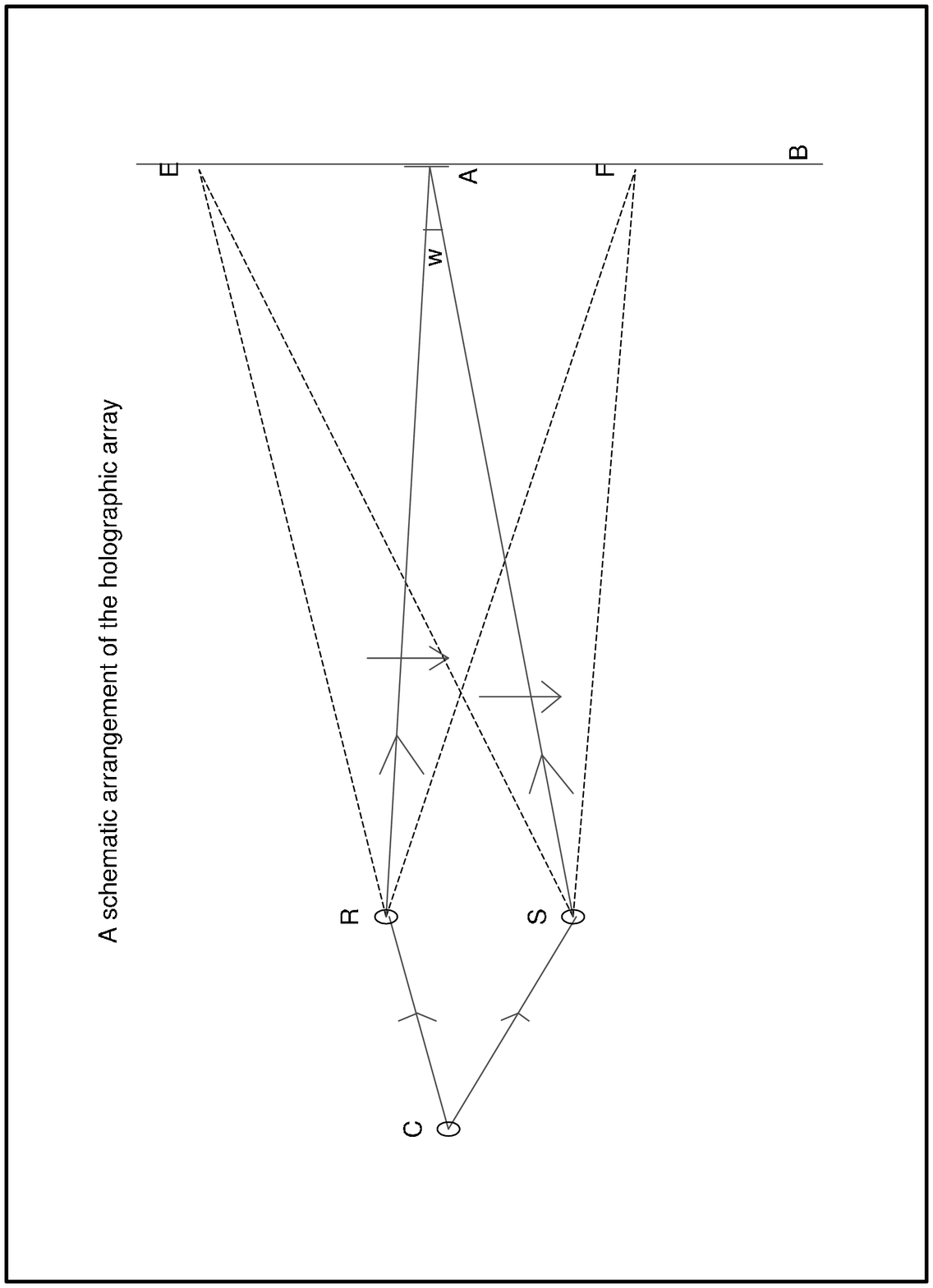}
\end{turn}}
     
     \caption{The constructive interference process of the subject and
reference waves over the small region $A$ shown in the middle of the figure. Other
small regions, denoted $E$ and $F$ are shown higher and lower than $A$. The lines,
representing the GW's 
which proceed to these regions are shown in dashed forms.  }
     \end{figure}

     \markright{CONCLUDING REMARKS}
 \protect \section{Concluding Remarks}  
 
 We have continued and expanded  our former discussion \cite{holo} regarding 
 the
 holographic properties of the GW's which, theoretically,  results from a 
 comparison between 
 optics holography 
  and the linearized version  of general relativity.
 It is argued  that the outcome of the gravitational holographic process, 
 which is the additional  curving of the relevant spacetime region 
 compared to neighbouring regions may be considered
 as if the spacetime geometry of the passing GW 
  was implanted upon this spacetime region. Thus, this finite spacetime region 
  which carry the geometry of the passing GW may be thought of as a surface
  formed by this GW which corresponds  
 to the trapped surfaces \cite{eppley,brill2,alcubierre} which are also; (1)  
 imprinted upon
 spacetime by passing GW's and (2) carry the same spacetime  geometry  as that
 of the generating GW's. Moreover, this correspondence is more emphasized by
 noting  that it has already been found \cite{tipler,yurtsever1} that 
  the collision between two plane GW's  results  in the overall 
 strenghtening of them \cite{tipler,yurtsever1} (corresponds to 
 constructive interference) 
 and 
  the  formation of a
 singularity, which is generally surrounded by a surface,  
 in the involved region. 
   \par  
   Now,  although these trapped  surfaces are generally discussed in the
   literature \cite{eppley,alcubierre,beig,abrahams} as formed from 
   strong GW's
    which do not conform to the linearized version of
   general relativity and the resulting weak GW's discussed here it should be
   noted, as mentioned,  that we only discuss here the situation 
    {\it in the presence} of  these
   GW's.   That is, as already emphasized in \cite{alcubierre}, weak GW's do not
   leave any impression upon the relevant spacetime region after passing and
   disappearing from it  but they obviously
   influence this finite region during their presence in it. Thus, although the
   GW's, including the weak ones, move with the velocity $c$ of light one may
   discuss either the situation at the very instant at which this wave passes
   this region or the situation at which a large number of similar waves are
   passing one after the other through this region.   Moreover, we
   discuss here the constructive interference of two GW's which have a larger
   influence  upon spacetime compared to that of the single one. And indeed it is
   shown here that these GW's does form {\it during their presence} in the relevant
   spacetime region a trapped surface which, as mentioned, remains and stays 
   so long as the forming waves does not disappear from it. \par
We note that unlike the EM holographic images which are 3-D surfaces seen by the
naked eye and so may be directly discussed in observational terms such
as length, distance, intensity etc the gravitational holographic images,
 actually, denote changes of
spacetime curvature and topology \cite{finkelstein,sorkin} 
which can not be directly 
observed and measured (see
discussion in \cite{holo}).  Moreover, the efforts to experimentally detect
\cite{ligo} GW's 
did not succeed thus far.  However,  this does not prevent us to  
 use  the equivalence principle 
\cite{mtw,bergmann} and  describe any physical event 
   in terms of either curved   
 spacetime in vacuum without resorting to any physical  interactions 
     or in terms of physical particles subjected 
to  physical correlations and forces in otherwise flat spacetime.  \par
In order to illustrate our meaning we refer to the linearized GW discussed here
and in \cite{holo} which changes spacetime curvature in the region passed by it
so that, as shown in
Figure 1,  $n$  test particles (TP) arrayed in circular form is transformed 
to an elliptic one. 
  This process may be discussed from the point of view of curved
spacetime so that the $n$ particles are represented by $n$ 
Einstein-Rosen bridges \cite{einstein} each of them is surrounded 
by an intrinsic trapped (minimal) 
surface \cite{eppley,brill2}.  And their being now in locations 
in which they are   generally closer to
each other  (as in an
ellipse array) is described by  an additional trapped surface \cite{brill2} 
(the $n+1$-st one)
which connect all the $n$
Einstein-Rosen bridges.    One may alternatively 
describe this process by saying that  $n$ physical 
particles, which are always situated in flat spacetime,  have undergone some
{\bf physical correlation} which  changes their initial spatial  array from    
 a circular form to an elliptic  one.

\begin{appendix}

\markright{APPENDIX A}

\section{APPENDIX A}

\protect \section{The linearized gravitational wave}

In this appendix we refer to  a TP which fall freely 
along the geodesic ${\cal A}$ and 
watches another TP falling freely along a neighbouring geodesic ${\cal B}$.
Referring to the 
general case of a coordinate system in which the basis $e_{\beta}$ changes
arbitrarily but smoothly from point to point and denoting the tangent  vector to
the geodesic $\hat{G(n,\tau)}$ as ${\bf u}=\frac{\partial \hat{G(n,\tau)}}{\partial \tau}$ 
one may write the velocity of
the TP along ${\cal B}$ relative to that along ${\cal A}$ as
$$ \nabla_{\bf u}{\bf n}=(n^{\beta};_{\gamma} u^{\gamma})e_{\beta}
\eqno (A1) $$   
$n^{\beta};_{\gamma}$ is the covariant derivative of $n^{\beta}$  given by 
\cite{mtw,bergmann}
$$
n^{\beta};_{\gamma}=\frac{dn^{\beta}}{dx^{\gamma}}+\Gamma^{\beta}_{\mu
\gamma}n^{\mu},  \eqno (A2) $$
where 
$$ \Gamma^{\beta}_{\mu \gamma}=g^{\nu \beta}\Gamma_{\nu \mu
\gamma}=\frac{1}{2}g^{\nu \beta}(g_{\nu \mu,\gamma}+g_{\nu \gamma,\nu}-g_{\mu
\gamma,\nu}) \eqno (A3) $$
The expression between the circular parentheses in $(A1)$ represents the
components of $\nabla_{\bf u}{\bf n}$ and is denoted by
$\frac{Dn^{\beta}}{d\tau}$. Thus, using Eq  $(A1)$ one may write \cite{mtw}
$$  
\frac{Dn^{\beta}}{d\tau}=n^{\beta};_{\gamma} u^{\gamma} 
=\frac{dn^{\beta}}{d\gamma}+\Gamma^{\beta}_{\mu
\gamma}n^{\mu}\frac{dx^{\gamma}}{d\tau} \eqno (A4)  $$
The acceleration of the TP along ${\cal B}$ relative to that along ${\cal A}$ is \cite{mtw} 
$$ \nabla_{\bf u}(\nabla_{\bf u}{\bf n})=-R, \eqno (A5)
$$ 
where $R$ is the Riemann curvature tensor given in component form as 
\cite{mtw,bergmann}
$$ R^{\alpha}_{\beta \gamma \delta}=\frac{\partial
\Gamma^{\alpha}_{\beta \delta}}{\partial x^{\gamma}}-\frac{\partial
\Gamma^{\alpha}_{\beta \gamma}}{\partial x^{\delta}}+\Gamma^{\alpha}_{\mu
\gamma}\Gamma^{\mu}_{\beta \delta}-\Gamma^{\alpha}_{\mu
\delta}\Gamma^{\mu}_{\beta \gamma} \eqno (A6) $$
Eq $(A5)$ may be written in component form as \cite{mtw} 
$$ \frac{D^2n^{\alpha}}{d\tau^2}=
-R^{\alpha}_{\beta \gamma \delta}u^{\beta}u^{\delta}n^{\gamma} \eqno (a7) 
$$
Now, returning to the local Lorentz frame  represented by the metric 
(\ref{e4}) one may realize that since, as mentioned, $x^0=\tau$ on the world
line $x^j=0$ of ${\cal A}$  the relation $(A7)$ reduces to the much simplified 
 form
$$ \frac{D^2n^j}{d\tau^2}=
-R^j_{0k0}n^k=-R_{j0k0}n^k  \eqno (A8) 
$$
We note that the transverse trace-free (TT) coordinate system may move, to first
order in the metric perturbation $h^{TT}_{jk}$, with TP ${\cal A}$ and with its proper
reference frame. Thus, to first order in $h^{TT}_{jk}$, one may identify the
time $t$ in the coordinate system $TT$ with the proper time $\tau$ of the 
TP ${\cal A}$
so as to have $R^{TT}_{j0k0}=R_{j0k0}$. In the last equality 
 the $R_{j0k0}$ at the right is
calculated in the proper reference frame of ${\cal A}$ and that at  the left is
calculated in the $TT$ coordinate system where it has been shown 
(see Eq (35.10) in \cite{mtw}) to  assume the very simple
form 
 of $$ R_{j0k0} = -\frac{1}{2}h^{TT}_{jk,00} \eqno (A9)
 $$ 
 One may notice that since the $TT$ coordinate system moves with the proper
 reference frame of TP ${\cal A}$ they are both denoted by the same  indices $(0,k,j)$ 
 without having to use primed and unprimed indices.    
Note that since the origin is attached to ${\cal A}$'s geodesic the components of the
separating vector ${\bf n}$ are, actually, the coordinates of ${\cal B}$. That is, 
denoting the coordinates of ${\cal A}$ and ${\cal B}$ by  $x^j_{{\cal A}}$ and 
$x^j_{{\cal B}}$ 
respectively  one may write $n^j=x^j_{{\cal B}}-x^j_{{\cal A}}=x^j_{{\cal
B}}-0=x^j_{{\cal B}}$. One may also notice
that at $x^j=0$ the connection coefficients $\Gamma^{\mu}_{\alpha \beta}$
satisfy $\Gamma^{\mu}_{\alpha \beta}=0$ for all $x^0$. This entails also the
vanishing of $\frac{d\Gamma^{\mu}_{\alpha \beta}}{d\tau}$ thereby the covariant
derivative $\frac{D^2n^j}{d\tau^2}$ becomes ordinary derivative and, using Eq
$(A9)$,   one may
write Eq $(A8)$ as $$ \frac{d^2x^j_{{\cal B}}}{d\tau^2}=
-R_{j0k0}x^k_{{\cal B}} =\frac{1}{2}(\frac{\partial ^2h^{TT}_{jk}}{\partial
t^2})x^k_{{\cal B}} 
\eqno  (A10)
$$
We choose the initial condition that the test particles at ${\cal A}$ and ${\cal
B}$ were at
rest before the wave arrives, that is, $x^j_{{\cal B}}=x^j_{{\cal B}(0)}$ when $h^{TT}_{jk}=0$. 
In this case the solution of Eq $(A10)$ is
$$ x^j_{{\cal B}}(\tau)=x_{{\cal B}(0)}^k(\delta_{jk}+\frac{1}{2}h^{TT}_{jk})_{at
{\cal A}}, \eqno (A11) $$ 
which represents the change in ${\cal B}$'s place due to the passing wave as calculated
in the proper reference frame of ${\cal A}$.  The $h^{TT}_{jk}$ at the right hand side 
of the last equation  represents the perturbation caused by 
advancing gravitational wave  which may be assumed to have any form. One may
choose any general waveform which may analytically expressed as an infinite
expansion of scalar, vector or tensor spherical harmonics \cite{thorne1} or one
may prefer an exact plane gravitational wave (see Section 35.9 in \cite{mtw}). 
     
\end{appendix}

\begin{appendix}

\markright{APPENDIX B}

\section{APPENDIX B}

\protect \section{The EM trapped surface}

As emphasized in \cite{mtw} (see Pages 961-962 there), 
 the EM plane wave equation {\it "has exactly the form of the equation for the
 gravitational plane wave"} and {\it "in the limit of very small wavelength the
 two solutions are completely indistinguishable. Their metrics are identical"
 etc}. Thus, in this limit one may use  the former process
 of finding the geometry of the  gravitational trapped surface 
 for calculating the corresponding  surface related to EM waves. We
 should only take into account that  an EM wave advancing along the $z$
 direction is characterized by the unit
 polarization vectors ${\bf e}_{\hat {\bf x}}, \ {\bf e}_{\hat {\bf y}}$. 
 Thus, as in Eq (\ref{e18}), one may write the EM metric 
 components
$ h^{EM}_{{\hat {\bf x}}{\hat{\bf x}}}, \ h^{EM}_{{\hat {\bf y}}{\hat{\bf y}}}$ 
as 
  $$ h^{EM}_{{\hat {\bf x}}{\hat{\bf x}}} =\Re 
\biggl( A_x{\bf e}_{{\hat {\bf x}}}e^{-ift}
e^{ik{\bf r}\hat{{\bf z}}} \biggr) = 
 A_x{\bf e}_{{\hat {\bf x}}} \cdot  
\cos(kz -ft)     \eqno{(B1)} $$  $$  h^{EM}_{{\hat {\bf y}}{\hat{\bf y}}}=\Re 
\biggl( A_y{\bf e}_{{\hat {\bf y}}}e^{-ift}
e^{ik{\bf r}\hat{{\bf z}}} \biggr) =
 A_y{\bf e}_{{\hat {\bf y}}} \cdot  
\cos(kz-ft),   $$
    where $A_x, \ A_y$ are the $x$ and $y$ amplitudes of the EM wave which are
    considered to be equal as the corresponding $xx$ and $yy$ components of the 
    amplitude  
    $A_+$ (see Pages (952-953) in \cite{mtw}). Using the
    last equation one may write the metric in the $({\hat{\bf x}},{\hat{\bf y}},
{\hat{\bf z}})$ coordinate system  as 
  $$  (ds^{EM})_{({\hat{\bf x}},{\hat{\bf y}},
{\hat{\bf z}})}^2=h^{EM}_{{\hat {\bf x}}{\hat{\bf x}}}dx^2+
h^{EM}_{{\hat {\bf y}}{\hat{\bf y}}}dy^2 =   
A_x\cos(kz -ft)\biggl({\bf e}_{{\hat {\bf x}}}dx^2+ 
{\bf e}_{{\hat {\bf y}}}dy^2\biggr) 
 \eqno{(B2)}  
    $$ 
 Using the  transformation relations 
 $x=\rho\cos(\phi), \ y=\rho\sin(\phi), \ z=z , \ \  
  {\bf e}_{\hat {\bf x}}=\cos(\phi){\bf e}_{\hat {\bf \rho}}-\sin(\phi){\bf
  e}_{\hat {\phi}}, \ \ {\bf e}_{\hat {\bf y}}=
  \sin(\phi){\bf e}_{\hat {\bf \rho}}+\cos(\phi){\bf
  e}_{\hat {\phi}}, \  \ {\bf e}_{\hat {\bf z}}= {\bf e}_{\hat {\bf z}},  
 $
 we   first express the  metric from Eq
$(B2)$   in the cylindrical coordinate system 
$({\hat{\bf \rho}}, \ {\hat{\bf \phi}}, \ {\hat {\bf z}})$ and 
use, as for the gravitational case (see the discussion after Eq
(\ref{e21})), the no-rotation assumption for which 
$h^{EM}_{{\hat {\bf \rho}}{\hat{\bf \phi}}}d\rho d\phi$ is identically zero 
  so that  
$$ (ds^{EM})_{({\hat{\bf \rho}},{\hat{\bf \phi}},
{\hat{\bf z}})}^2=
h^{EM}_{{\hat {\bf \rho}}{\hat{\bf \rho}}}d\rho^2+
h^{EM}_{{\hat {\bf \phi}}{\hat{\bf \phi}}}d\phi^2 
=   
A_{\rho,\phi}\cos(kz -ft)\cdot $$
$$ \cdot \biggl[\biggl\{\left(\cos^3(\phi)+\sin^3(\phi)\right){\bf e}_{{\hat{\bf
\rho}}}+\left(\sin^2(\phi)\cos(\phi)-\cos^2(\phi)\sin(\phi)\right)
{\bf e}_{{\hat{\bf \phi}}}\biggr\}d\rho^2 +\eqno{(B3)} $$
 $$ +\biggl\{\left(\sin^2(\phi)\cos(\phi)+\cos^2(\phi)\sin(\phi)\right){\bf e}_{{\hat{\bf \rho}}}
+\left(\cos^3(\phi)-\sin^3(\phi)\right){\bf e}_{{\hat{\bf
\phi}}}\biggr\}\rho^2d\phi^2 \biggr]
$$
 Using again, as in Eq (\ref{e23}), the relations 
$ x=a(\rho)\cos(\phi), \ \ y=a(\rho)\sin(\phi), \ \ z=b(\rho) $ 
  one may write the metric  as 
 $$  
 (ds^{EM})^2=dx^2+dy^2+dz^2=(a^2(\rho)_{\rho}+b^2(\rho)_{\rho})d\rho^2+
 a^2(\rho)d\phi^2=  $$   $$ = 
 h^{EM}_{{\hat {\bf \rho}}{\hat{\bf \rho}}}d^2\rho +
h^{EM}_{{\hat {\bf \phi}}{\hat{\bf \phi}}}d^2\phi  = 
\cos(kz -ft)\cdot \eqno{(B4)}  $$  $$ \cdot \biggl[\biggl\{\left(\cos^3(\phi)+\sin^3(\phi)\right){\bf e}_{{\hat{\bf
\rho}}}+\left(\sin^2(\phi)\cos(\phi)-\cos^2(\phi)\sin(\phi)\right)
{\bf e}_{{\hat{\bf \phi}}}\biggr\}d\rho^2 +  $$ 
  $$+\biggl\{\left(\sin^2(\phi)\cos(\phi)+\cos^2(\phi)\sin(\phi)\right){\bf e}_{{\hat{\bf \rho}}}
+\left(\cos^3(\phi)-\sin^3(\phi)\right){\bf e}_{{\hat{\bf
\phi}}}\biggr\}\rho^2d\phi^2 \biggr]
$$
The appropriate expressions $a(\rho), \ a(\rho)_{\rho}, \ b(\rho)$ which
determine the geometry of the relevant trapped surface are obtained from Eq
(B4) as
  $$ a(\rho)= \rho\biggl[\cos(kz -ft)\biggl\{\left(\sin^2(\phi)\cos(\phi)+\cos^2(\phi)\sin(\phi)\right){\bf e}_{{\hat{\bf \rho}}}
+\left(\cos^3(\phi)-\sin^3(\phi)\right){\bf e}_{{\hat{\bf
\phi}}}\biggr\}  \biggr]^{\frac{1}{2}} $$ 
$$ a(\rho)_{\rho}= \biggl[\cos(kz -ft)\biggl\{\left(\sin^2(\phi)\cos(\phi)+\cos^2(\phi)\sin(\phi)\right){\bf e}_{{\hat{\bf \rho}}}
+\left(\cos^3(\phi)-\sin^3(\phi)\right){\bf e}_{{\hat{\bf
\phi}}}\biggr\} \biggl]^{\frac{1}{2}}  $$ 
$$ b(\rho)=\int d\rho\biggl[\cos(kz -ft) \biggl\{\left(\cos^3(\phi)+\sin^3(\phi)\right){\bf e}_{{\hat{\bf
\rho}}}+\left(\sin^2(\phi)\cos(\phi)-\cos^2(\phi)\sin(\phi)\right)
{\bf e}_{{\hat{\bf \phi}}}\biggr\} - $$   $\ \ \ \ \ \ \ \ \ \ \ \ \ -a^2(\rho)_{\rho}\biggr]^{\frac{1}{2}}
\ \ \ \ \ \ \ \ \ \ \ \ \ \ \ \ \ \ \ \ \ \ \ \ \ \ \ \ \ \ \ \ \ \ \ \ \ \ \ \
\ \ \ \ \ \ \ \ \ \ \ \ \ \ \ \ \ \ \ \ \ \ \  \ \ \ \ \ \ \ \ \ \ \ \ \ \ \ \ \
\ \ \ \ \ \ \ \ \ \ (B5)$    
  
\end{appendix}

 \markright{TABLE 1}
  
 \begin{table}
\caption{\label{table1} The table compares the holographic evolutions for the
electromagnetic and gravitational waves from the initial stage 
of following the separate subject and reference
waves to the final stage of reconstructing the subject wave and forming the
trapped surfaces.}
     \begin{center}
      \begin{tabular}{|l|l|l|l|l|} 
        \  N&The \ holographic \ evolutions   &The \ electromagnetic &
	The \ gravitational \\ & for \ the \ electromagnetic 
         & wave \  holography. & wave \ holography. \\ 
	 & and \ gravitational \ waves. & & \\
        \hline \hline
$1 $ &The \ initial \ subject \ and \ reference  & $ \scriptstyle{
S^{EM}_{jk}=(A_{S_j}{\bf e}_{S_j}+A_{S_k}{\bf e}_{S_k}) \cdot } 
 $&$ \scriptstyle{ h^{TT}_{S_{jk}}=
(A_{+S}{\bf e}_{+_{S_{jk}}}+
A_{\times S}{\bf e}_{\times_{S_{jk}}}) \cdot }$ \\ 
$$ &  waves \ where \ ${\bf e}_{S_j}$,   \ ${\bf e}_{S_k}$, \ ${\bf e}_{R_j}$, \
${\bf e}_{R_k}$   
   &$ \scriptstyle{ \cdot  e^{-ift}e^{ik{\bf r}_S{\hat {\bf n}_S}} }$& 
$ \scriptstyle{ \cdot e^{-ift}e^{ik{\bf
r}_S\hat{{\bf n}_S}} } $  \\
 $  $& are \ the \ EM \ polarization \ vectors  
   && \\ $$  &  and \ ${\bf e}_{+_{S_{jk}}}$, \ 
${\bf e}_{+_{S_{jk}}}$, \  ${\bf e}_{\times_{R_{jk}}}$,  \ 
${\bf e}_{+_{R_{jk}}}$ 
\ are       &  
  $ \scriptstyle{ R^{EM}_{jk}=(A_{R_j}{\bf e}_{R_j}+A_{R_k}{\bf e}_{R_k}) \cdot }  
 $  &$ \scriptstyle{  h^{TT}_{R_{ jk}}=  (A_{+R}{\bf e}_{+_{R_{jk}}}+
A_{\times R}{\bf e}_{\times_{R_{jk}}}) \cdot  } $ \\ 
$$ &the \  GW 
\ polarization \ tensors.     & 
$ \scriptstyle{\cdot  e^{-ift}e^{ik{\bf r}_R{\hat {\bf n}_R}} }$&$ 
 \scriptstyle{ \cdot e^{-ift}e^{ik{\bf
r}_R\hat{{\bf n}_R}} } $ \\ 
\hline 
$ 2 $ & The  \ total \ intensities \ of \ the \ inter-   
&$\scriptstyle{ I^{EM}_{S+R}= I^{EM}_{(S+R)_0}+I^{EM}_{(S+R)_1}= }$ & $ 
\scriptstyle{ I^{TT}_{S+R}= I^{TT}_{(S+R)_0} +I^{TT}_{(S+R)_1}= }$ \\ 
$$ & -fering \ subject \ and \ reference \ waves      &$ 
\scriptstyle{ \biggl( A_{S_j}{\bf e}_{S_j}+A_{S_k}{\bf e}_{S_k} \biggr) \cdot } $&$ 
\scriptstyle{ \biggl( A_{+_S}{\bf e}_{+_{S_{jk}}}
 + A_{\times_S}{\bf e}_{\times_{S_{jk}}} 
  \biggr) \cdot }$ \\ 
$  $ &  where \ $I^{EM}_{(S+R)_0}$,  \ $I^{TT}_{(S+R)_0}$ \ denote \ the &$
 \scriptstyle{ \cdot \biggl(  A_{S_j}{\bf e}_{S_j}+A_{S_k}{\bf e}_{S_k} \biggr)^* +   }    $&  
$ \scriptstyle{ \cdot \biggl(  A_{+_S}{\bf e}_{+_{S_{jk}}}
 + A_{\times_S}{\bf e}_{\times_{S_{jk}}} \biggr)^* +}  $ \\ 
$  $ &terms \ which \ do \ not \ depend \ upon   &$ 
\scriptstyle{ + \biggl( A_{R_j}{\bf e}_{R_j}   +A_{R_k}{\bf e}_{R_k} \biggr) \cdot }    
  $&  $ \scriptstyle{ + 
 \biggl( A_{+_R}{\bf e}_{+_{R_{jk}}} 
 + A_{\times_R}{\bf e}_{\times_{R_{jk}}} \biggr) \cdot }
   $ \\ 
$  $ & the \  complex \ exponentials \ and    &$ \scriptstyle{
\cdot \biggl( A_{R_j}{\bf e}_{R_j}   +A_{R_k}{\bf e}_{R_k} \biggr)^* +  }   $&  
$ \scriptstyle{  \cdot \biggl(  A_{+_R}{\bf e}_{+_{R_{jk}}} 
 + A_{\times_R}{\bf e}_{\times_{R_{jk}}} \biggr)^*  +}
   $ \\ 
$  $ & $I^{EM}_{(S+R)_1}$, \   $I^{TT}_{(S+R)_1}$ 
\ denote  \ the \ terms      &$ \scriptstyle{ + \biggl[ \biggl( 
 A_{S_j}{\bf e}_{S_j}  
+ A_{S_k}{\bf e}_{S_k} \biggr) \cdot } $&  $ \scriptstyle{+ \biggl[ \biggl( 
 A_{+_S}{\bf e}_{+_{S_{jk}}}  
 + A_{\times_S}{\bf e}_{\times_{S_{jk}}} \biggr)  \cdot }
 $\\
 $ $ &  which \  depend \ upon \ them &$ 
  \scriptstyle{ \cdot \biggl( A_{R_j}{\bf e}_{R_j}+ A_{R_k}{\bf e}_{R_k}
  \biggr) \biggr] \cdot  }$ & $ 
  \scriptstyle{ \cdot \biggl(  A_{+_R}{\bf e}_{+_{R_{jk}}} 
 + A_{\times_R}{\bf e}_{\times_{R_{jk}}} \biggr) \biggr] \cdot } $ \\ 
 $  $ &$      $&$ \scriptstyle{ \cdot e^{ik({\bf r}_S{\bf n}_S-{\bf r}_R{\bf
 n}_R)} 
  } $ 
 &  $ \scriptstyle{ \cdot 
 e^{ik({\bf r}_S{\bf n}_S-{\bf r}_R{\bf n}_R)} }$ \\ 
\hline
$3$ & The \ exposure \ related \ to  \ the \ total   
&$\scriptstyle{  E^{EM}=E^{EM}_0+ E^{EM}_1=I^{EM}_{S+R}\tau_e=  } $& $ 
    \scriptstyle{  E^{TT}=E^{TT}_0+ E^{TT}_1 = I^{TT}_{S+R}\tau_e= }   $ \\
$$ & intensities \ where \ $\tau_e $ \ is \ the \ expo-   &$
\scriptstyle{  =I^{EM}_{(S+R)_0}\tau_e  + I^{EM}_{(S+R)_1}\tau_e } $ & $  
  \scriptstyle{ =I^{TT}_{(S+R)_0}\tau_e  + I^{TT}_{(S+R)_1}\tau_e } $  \\ 
  $$&-sure \ time &$$&$$ \\
\hline
$4  $ & The  \ optics \ and \ GW \ holograms      &$ \scriptstyle{
 {\bf t}_{E^{EM}}={\bf t}^{EM}(E^{EM}_0)+ E^{EM}_1 \frac{d{\bf
t}_E}{dE}|_{E^{EM}_0} }    $&$ \scriptstyle{
 {\bf t}_{E^{TT}}={\bf t}^{TT}(E^{TT}_0)+ E^{TT}_1 \frac{d{\bf
t}_E}{dE}|_{E^{TT}_0}} 
    $  \\
$ $ & expressed \ by \ the \ respective    & $  \scriptstyle{ 
+ \frac{1}{2}(E^{TT}_1 )^2\frac{d^2{\bf t}_E}{dE^2}|_{E^{TT}_0}+\cdots } $ & $ \scriptstyle{ 
+ \frac{1}{2}(E^{TT}_1 )^2\frac{d^2{\bf t}_E}{dE^2}|_{E^{TT}_0}+\cdots } $ \\   
$$ & amplitude \ transmittances \ ${\bf t}_{E^{EM}}$    
    &$  $&$  $  \\
$$ &and  \ ${\bf t}_{E^{TT}}$ \ as \ functions \ of   \ the 
  &$$&$$ \\
$$ &exposures \ $E^{EM}$,  \  $E^{TT}$ \  in  \ the
 & $ $&$ $ \\
 $$ & limit \ of \ linear \ recording &$$&$$ \\ 
\hline 
$5$ &For \  reconstructing \ the \ holographic  &$ 
\scriptstyle{ R^{EM}_{jk}=(A_{R_j}{\bf e}_{R_j}+A_{R_k}{\bf e}_{R_k}) \cdot } 
$&$\scriptstyle{  h^{TT}_{R_{
jk}}=  (A_{+_R}{\bf e}_{+_{R_{jk}}}+
A_{\times_R}{\bf e}_{\times_{R_{jk}}}) \cdot  } $ \\ 
$$ & EM \ and \ gravitational \ images \ the  &$  
\scriptstyle{\cdot  e^{-ift}e^{ik{\bf r}_R{\hat {\bf n}_R}} } 
 $&$  
 \scriptstyle{ \cdot e^{-ift}e^{ik{\bf
r}_R\hat{{\bf n}_R}} } $ \\ 
$$ & reference \ waves \ $R^{EM}_{jk}, \ h^{TT}_{R_{
jk}}$ \ are   &$    $&$    $ \\ 
$$ &sent \  to \ the \  respective \ holograms 
&&\\
$$& ${\bf t}_{E^{EM}}$ \ and  
\ ${\bf t}_{E^{TT}}$ &$ $&$ $ \\
\hline
$6$ & The \ reference \ waves \ $R^{EM}_{jk}$ \ and \ $h^{TT}_{R_{
jk}}$   &$ \scriptstyle{ R^{EM}_{jk}\cdot {\bf t}_{E^{EM}}=C_{EM}\cdot
S^{EM}_{jk} =} $  &$ 
\scriptstyle{  h^{TT}_{R_{
jk}}\cdot {\bf t}_{E^{TT}}=C_{TT}\cdot h^{TT}_{S_{jk}}= }  $ \\
$$ &are \  decomposed \ at \ the \ respective  &$\scriptstyle{ =C_{EM} 
\cdot  (A_{S_j}{\bf e}_{S_j}+A_{S_k}{\bf e}_{S_k}) \cdot }
$ &$
\scriptstyle{ = C_{TT}\cdot (A_{+S}{\bf e}_{+_{S_{jk}}}+
A_{\times_S}{\bf e}_{\times_{S_{jk}}}) \cdot } $ \\ 
$$ & holograms \ ${\bf t}_{E^{EM}}$ \ and  
\ ${\bf t}_{E^{TT}}$ \ in \ the  
 &$ \scriptstyle{ \cdot  e^{-ift}e^{ik{\bf r}_S{\hat {\bf n}_S}} }$    
  &$  \scriptstyle{ \cdot e^{-ift}e^{ik{\bf
r}_S\hat{{\bf n}_S}} }    $ \\ 
$$& limit \  of \ linear \ recording \ to \ the   &$$ &$$ \\ 
$$& holographic \  images \  $S^{EM}_{jk}$ \ and \  $h^{TT}_{S_{
jk}}  $  &$$&$$ \\ $$ & where \ $C_{EM}$ \ and $C_{TT}$ \ are \ constants&$    
$&$    $ \\
\end{tabular} 
\end{center}
\end{table}

 \begin{table}
 \begin{center} 
 \begin{tabular}{|l|l|l|l|l|} 
 
 \  N&The \ holographic \ evolutions   &The \ electromagnetic &
	The \ gravitational \\ & for \ the \ electromagnetic 
         & wave \  holography.    & wave \ holography.   \\ 
	 & and \ gravitational \ waves. & &   \\ 
	& {\it continued} & {\it continued} & {\it continued} \\
        \hline \hline

$7$&The \ geometry \ of \ the \ trap- & $\scriptstyle{ a^{EM}(\rho)=
 \rho\biggl[\cos(kz -ft)\biggl\{\biggl(\sin^2(\phi)\cos(\phi)+ }
   $&$ \scriptstyle{  a^{TT}(\rho)= 
\rho\biggl[ \cos(kz -ft)\biggl\{ 
 \frac{\sin(4\phi)}{2} \cdot } $ \\ 
$$ &-ped \ surfaces \ resulting \ from &$\scriptstyle{ +\cos^2(\phi)\sin(\phi)\biggr)
{\bf e}_{{\hat{\bf \rho}}} 
 +\biggl(\cos^3(\phi)- } $&
 $\scriptstyle{ \cdot (A_+-A_{\times})\biggl( {\bf e}_{\hat {\bf \rho}}\otimes 
{\bf e}_{\hat {\bf \phi}} + {\bf e}_{\hat {\bf \phi}}\otimes 
{\bf e}_{\hat {\bf \rho}}\biggr)+   } $ \\
$$& the \  influences \ of \ the \ EM $$&$\scriptstyle{ -\sin^3(\phi)\biggr){\bf e}_{{\hat{\bf
\phi}}}\biggr\}  \biggr]^{\frac{1}{2}} }$ &$ \scriptstyle{ +\biggl(
A_+\cos^2(2\phi)+A_{\times}\sin^2(2\phi)\biggr) \cdot }$ \\ 
$$ &and \  gravitational \ waves. & &$ \scriptstyle{ \cdot \biggl({\bf e}_{\hat {\bf \phi}}\otimes 
{\bf e}_{\hat {\bf \phi}}-{\bf e}_{\hat {\bf \rho}}\otimes 
{\bf e}_{\hat {\bf \rho}}\biggr) \biggr\}\biggr]^{\frac{1}{2}} } $\\
$$& The \ trapped \ surface \ for     && \\
$$& the \  EM \ case \   results \ for &$
\scriptstyle{ a^{EM}(\rho)_{\rho}= 
\biggl[\cos(kz -ft)\biggl\{\biggl(\sin^2(\phi)\cos(\phi)+ }
 $&
$\scriptstyle{ a^{TT}(\rho)_{\rho}= \biggl[\cos(kz -ft)\biggl\{ 
 \frac{\sin(4\phi)}{2}  \cdot } $ \\ 
$$& the \  special \ condition \  of   $$&$\scriptstyle{ + \cos^2(\phi)\sin(\phi)\biggr){\bf e}_{{\hat{\bf \rho}}}
+ \biggl(\cos^3(\phi)-    } $&
$\scriptstyle{ \cdot (A_+-A_{\times})\biggl( {\bf e}_{\hat {\bf \rho}}\otimes 
{\bf e}_{\hat {\bf \phi}} + {\bf e}_{\hat {\bf \phi}}\otimes 
{\bf e}_{\hat {\bf \rho}}\biggr)+   } $ \\
$$& very \ small \ wavelength.    &$\scriptstyle{ -\sin^3(\phi)\biggr){\bf e}_{{\hat{\bf
\phi}}}\biggr\} \biggl]^{\frac{1}{2}} } $&$ \scriptstyle{ +\biggl(
A_+\cos^2(2\phi)+A_{\times}\sin^2(2\phi)\biggr) \cdot }  $\\
$$& Note \  that \  the \ geometries &&$\scriptstyle{ \cdot \biggl({\bf e}_{\hat {\bf \phi}}\otimes 
{\bf e}_{\hat {\bf \phi}}-{\bf e}_{\hat {\bf \rho}}\otimes 
{\bf e}_{\hat {\bf \rho}}\biggr) \biggr\}\biggr]^{\frac{1}{2}} } $ \\
$$& given \ by \ $a^{EM}(\rho),\ a^{EM}(\rho)_{\rho}$ && \\
$$& $ b^{EM}(\rho), \ a^{TT}(\rho), \ a^{TT}(\rho)_{\rho}, 
 $       
&$ \scriptstyle{ b^{EM}(\rho)=\int d\rho\biggl[\cos(kz -ft) 
\biggl\{ \biggl(\cos^3(\phi)+    } $&
$\scriptstyle{ b^{TT}(\rho)=\int d\rho\biggl[\cos(kz -ft)
\biggl\{\frac{\sin(4\phi)}{2} }  $ \\ 
$$& $b^{TT}(\rho)$ \ are \ for \ the  
 &$\scriptstyle{ +\sin^3(\phi)\biggr){\bf e}_{{\hat{\bf
\rho}}}+ \biggl(\sin^2(\phi)\cos(\phi)- }$&
$\scriptstyle{ \cdot (A_{\times}-A_+)\biggl( {\bf e}_{\hat {\bf \rho}}\otimes 
{\bf e}_{\hat {\bf \phi}} + {\bf e}_{\hat {\bf \phi}}\otimes 
{\bf e}_{\hat {\bf \rho}}\biggr)+   } $ 
\\
$$ & embedding \  surfaces \ only       
 &$\scriptstyle{-\cos^2(\phi)\sin(\phi)\biggr)
{\bf e}_{{\hat{\bf \phi}}}\biggr\} - 
(a^{EM}(\rho)_{\rho})^2\biggr]^{\frac{1}{2}} }  
$&$ 
 \scriptstyle{ +\biggl(
A_+\cos^2(2\phi)+A_{\times}\sin^2(2\phi)\biggr) \cdot } $ \\
$$& (see \ text).   
 &   & $
\scriptstyle{ \cdot \biggl({\bf e}_{\hat {\bf \rho}}\otimes 
{\bf e}_{\hat {\bf \rho}}-{\bf e}_{\hat {\bf \phi}}\otimes 
{\bf e}_{\hat {\bf \phi}}\biggr) \biggr\}
 -(a^{TT}(\rho)_{\rho})^2\biggr]^{\frac{1}{2}} } $
\\

 \end{tabular} 
\end{center}
\end{table}

\newpage

\markright{REFERENCES}

 \bigskip \bibliographystyle{plain}

\begin{thebibliography}{99}
  
      \bigskip \parindent 1 in 


\bibitem{mtw}  {\rm "Gravitation"}, C. W. Misner, K. S. Thorne and J. A.
Wheeler, Freeman, San Francisco (1973)
 \bibitem{thorne1} K. S. Thorne, {\rm ``Multipole expansions of gravitational
 radiation''}, Rev. Mod. Phys {\bf 52}, 299 (1980)
\bibitem{thorne2}  K. S. Thorne, {\rm ``Gravitational wave research: Current
status and future prospect''},  Rev. Mod. Phys {\bf 52}, 285 (1980)
\bibitem{brill1} D. Brill and R. W. Lindquist, {\rm ``Interaction energy in
Geometrodynamics''}, Phys. Rev, {\bf 131}, 471-476 (1963)
\bibitem{eppley} K. Eppley, {\rm ``Evolution of time-symmetric gravitational
waves: Initial data and apparent horizons''}, 
Phys. Rev D, {\bf 16}, 1609 (1977); 
\bibitem{tipler} F. J. Tipler, {\rm `` Singularities from colliding plane
gravitational waves  ''}, Phys. Rev D, {\bf 22}, 2929 (1980)

\bibitem{yurtsever1} U. Urtsever, {\rm ``Singularities in the collisions of
almost-plane gravitational waves''}, Phys. Rev D, {\bf 38}, 1731 (1988); {\rm
Colliding almost-plane gravitational waves: Colliding plane waves and general
properties of almost-plane wave spacetimes''}, Phys. Rev D, {\bf 37}, 2803
(1988)
\bibitem{beig} R. Beig and N. O. Murchadha, {\rm `` Trapped surfaces due to
concentration of gravitational radiation''}, Phys. Rev Lett, {\bf 66}, 2421 (1991)

\bibitem{abrahams} A. M. Abrahams and C. R. Evans, {\rm ``Trapping a geon: Black
hole formation by an imploding gravitational wave''},  Phys. Rev D, {\bf 46}, R4117
(1992)
\bibitem{alcubierre} M. Alcubierre, G. Allen, B. Brugmann, G. Lanfermann, E.
Seidel, W. M. Suen and M. Tobias, {\rm ``Gravitational collapse of gravitational
waves in 3-D numerical relativity''}, Phys. Rev D, {\bf 61}, 041501 (2000); A. P. Gentle, D. E. Holz and W. A. Miller, {\rm ``Apparent horizons
in simplical Brill wave initial data"}, gr-qc/9812057; A. P. Gentle, 
{\rm "Simplical
Brill wave initial data"},  gr-qc/9901071; S. M. Miyama, {\rm "Time evolution of pure gravitational wave"} 
Prog. Theor. Phys, {\bf 65}, 894 (1981)
\bibitem{brill2}  D. R. Brill, {\rm ``On the positive definite mass of the
Bondi-Weber-Wheeler time-symmetric gravitational waves''}, Ann. Phys, {\bf 7}, 466 (1959); D. R. Brill,
Suppl. Nuovo. Cimento, {\bf 2}, 1-56 (1964); D. R. Brill and J. B. Hartle, {\rm
``Method of the self-consistent field in General Relativity and its application
to the gravitational geon''}, Phys.
Rev B, {\bf 135}, 271 (1964) 
\bibitem{adm}  R. Arnowitt, S. Desser and C. W. Misner, {\rm ``The dynamics of General
Relativity''} In {\rm ``Gravitation: An Introduction to current research''}, ed.
L. Witten, Wiley, New-York (1962)
\bibitem{nakamura} T. Nakamura,  {\rm "General solutions of the linearized Einstein
equations and initial data for 3-D time evolution of pure gravitational waves"}, 
 Prog. Theor. Phys, {\bf 72}, 746 (1984)
\bibitem{anninos} P. Anninos, J. Masso, E. Seidel, W. M. Suen and M. Tobias,
{\rm ``Dynamics of gravitational waves in 3D: Formulations, methods and
tests''}, Phys. Rev D, {\bf 56}, 842 (1997)
\bibitem{hawking}  S. W. Hawking and G. F. R. Ellis, {\rm ``The large scale
structure of spacetime''}, Cambridge, London (1973)
\bibitem{holo}  D. Bar, {\rm``The gravitational wave holography''},  gr-qc/0509052, to be published in IJTP
\bibitem{gabor} D. Gabor, {\bf ``A new microscopic Principle''}, 
Nature, {\bf 161}, 777 (1948); {\rm `` Microscopy by reconstructed
wavefronts''}, Proc. Roy. Soc, {\bf
A197}, 454 (1949);  {\rm `` Microscopy by reconstructed
wavefronts: II''},    Pro. Roy. Soc, {\bf B64}, 449 (1951)
\bibitem{collier} R. J. Collier, C. B. Burckhardt and L. H. Lin, {\rm ``Optical
Holography''}, Academic Press (1971) 
\bibitem{kuchar1}  K. Kuchar, {\rm ``Ground state functional of the linearized
gravitational field''}, J. Math. Phys, {\bf 11}, 3322 (1970)
\bibitem{kuchar2}  K. Kuchar,  {\rm ``Canonical quantization of cylindrical
gravitational waves''}, Phys. Rev D , {\bf 4}, 955  (1971)
\bibitem{bergmann}  P. G. Bergmann,   {\rm ``Introduction to the theory of
Relativity''}, Dover, New York (1976)
\bibitem{bernstein} D. Bernstein, D. Hobill, E. Seidel and L. Smarr, {\rm
``Initial data for the black hole plus Brill wave spacetime''}, Phys. Rev
D, {\bf 50}, 3760 (1994)
\bibitem{urtsever2} U. Urtsever, {\rm `` Quantum field theory in a colliding
plane-wave background''}, Phys. Rev D, {\bf 40}, 360 (1989) 

 
\bibitem{einstein}  A. Einstein and N. Rosen, {\rm ``The particle problem in the
general theory of relativity''}, Phys. Rev, {\bf 48}, 73 (1935)
 \bibitem{finkelstein} D. Finkelstein and E. Rodriguez, {\rm ``Relativity of topology
 and dynamics''}, Inter, J, Theor, Phys, {\bf 23}, 1065-1098 (1984)
 \bibitem{sorkin} R. D. Sorkin, {\rm ``Topology change and monopole creation''},
 Phys. Rev D, {\bf 33}, 978 (1986) 
\bibitem{spiegel} M. R. Spiegel, {\rm ``Vector analysis''}, Schaum's Outline
Series,McGraw-hill, New-York (1959) 
\bibitem{ligo}  B. Abbott {\it et al}, {\rm ``Analysis of  first LIGO science
data for stochastic gravitational waves''}, Phys. Rev D, {\bf 69}, 122004 (2004);  F. Acernese {\it et al}, {\rm "Status of VIRGO"}, Class, Quantum Grav, 
{\bf 19}, 1421 (2002);  K. Danzmann, {\rm "GEO-600 a 600-m laser interferometric gravitational 
wave antenna"}, In {\rm "First Edoardo Amaldi conference on gravitational wave 
experiments"}, E. Coccia, G. Pizella and F. Ronga, eds, World Scientific, Singapore
(1995);  M. Ando and the TAMA collaboration, {\rm "Current status of TAMA"}, 
Class. Quantum Grav, {\bf 19}, 1409 (2002)







\end{thebibliography}

\end{document}